\documentclass[12pt]{article}

\usepackage{geometry} 	
\usepackage{setspace}   	
\usepackage{amsmath,amssymb,amsfonts,amsthm}		
\usepackage[T1]{fontenc}
\usepackage{libertinus}
\usepackage{libertinust1math}
\usepackage{bm}                                 

  \everydisplay{%
    \abovedisplayskip=10pt%
    \belowdisplayskip=10pt%
    \abovedisplayshortskip=6pt%
    \belowdisplayshortskip=6pt%
  }
\usepackage{enumitem}

\usepackage{graphicx}
\usepackage{subcaption}

\usepackage{fancyhdr}			

\usepackage{xcolor}				
\usepackage{soul} 				

\usepackage{hyperref}			
	\hypersetup{colorlinks=true, citecolor = black, linkcolor = black, urlcolor=black}

\usepackage{tocloft}			

\usepackage{natbib}
\usepackage[USenglish]{babel}
\usepackage[style=iso]{datetime2} 

\usepackage[capitalise]{cleveref}

\usepackage{dutchcal} 

\usepackage[textsize=tiny,disable]{todonotes} 
\usepackage{marginnote}
\newcommand{\mytodo}[2][]{{
 \let\marginpar\marginnote
 \reversemarginpar
 \todo[#1]{#2}}}

\usepackage{booktabs} 
\usepackage{multicol}
\usepackage{multirow}

\usepackage{array}
\newcolumntype{N}{c@{}S}
\newcolumntype{H}{>{\setbox0=\hbox\bgroup}c<{\egroup}@{}}
\newcolumntype{L}[1]{>{\raggedright\let\newline\\\arraybackslash\hspace{0pt}}m{#1}}
\newcolumntype{C}[1]{>{\centering\let\newline\\\arraybackslash\hspace{0pt}}m{#1}}
\newcolumntype{R}[1]{>{\raggedleft\let\newline\\\arraybackslash\hspace{0pt}}m{#1}}

\usepackage[round-mode=places, table-number-alignment=center]{siunitx}

\usepackage{lscape} 

\usepackage[raggedright]{titlesec} 
\makeatletter
\renewcommand*{\@seccntformat}[1]{\csname the#1\endcsname\hspace{20pt}} 
\makeatother

\titleformat*{\section}{\fontsize{16}{16}\bfseries}

\titlespacing*{\paragraph}
  {0pt}
  {0.25em}
  {0.5em}

\titlespacing*{\section}
  {0pt}
  {1em}
  {0.5em}

\titlespacing*{\subsection}
  {0pt}
  {0.5em}
  {0.25em}

\titlespacing*{\subsubsection}
  {0pt}
  {0.25em}
  {0.125em}

\definecolor{darkpowderblue}{rgb}{0.0, 0.2, 0.6}
\definecolor{carmine}{rgb}{0.59, 0.0, 0.09}
\definecolor{cadmiumgreen}{rgb}{0.0, 0.42, 0.24}
\definecolor{burntorange}{rgb}{0.8, 0.33, 0.0}

\newtheorem{corollary}{Corollary}

\newtheorem{proposition}{Proposition}

\newtheorem{lemma}{Lemma}
\newtheorem{algo}{Algorithm}

\newtheorem{conjecture}{Conjecture}
    \crefname{conjecture}{Conjecture}{Conjectures}

\makeatletter
\newtheorem*{rep@theorem}{\rep@title}
\newcommand{\newreptheorem}[2]{%
\newenvironment{rep#1}[1]{%
 \def\rep@title{#2 \ref{##1}}%
 \begin{rep@theorem}}%
 {\end{rep@theorem}}}
\makeatother

\newreptheorem{theorem}{Theorem}
\newreptheorem{proposition}{Proposition}
\newreptheorem{lemma}{Lemma}

\newcommand{\myproposition}[3]{
\noindent 
\begin{proposition}[#1] \label{#2} \hspace*{0cm}\\
	#3
\end{proposition}
}

\newcommand{\myrepproposition}[3]{
\noindent 
\begin{repproposition}{#2}[#1] \hspace*{0cm}\\
	#3
\end{repproposition}
}

\newcommand{\propositionmarker}[1]{%
  \refstepcounter{proposition}%
  \label{#1}%
}

\newcounter{propositionpart}[proposition]

\makeatletter
\renewcommand{\p@propositionpart}{\theproposition}

\makeatother
\crefalias{propositionpart}{proposition}

\newcommand{\propositionpartmarker}[1]{%
  \refstepcounter{propositionpart}%
  \label{#1}%
}

\newcommand{\mylemma}[3]{
\noindent 
\begin{lemma}[#1] \label{#2} \hspace*{0cm}\\
	#3
\end{lemma}
}

\newcommand{\myreplemma}[3]{
\noindent 
\begin{replemma}{#2}[#1] \hspace*{0cm}\\
	#3
\end{replemma}
}

\newcommand{\lemmamarker}[1]{%
  \refstepcounter{lemma}%
  \label{#1}%
}

\newcommand{\myconjecture}[3]{
\noindent 
    \begin{conjecture}[#1] \label{#2} \hspace*{0cm}\\
      #3
    \end{conjecture}
}

\newcommand{\myalgo}[3]{
\noindent 
\begin{algo}[#1] \label{#2} \hspace*{0cm}\\
	#3
\end{algo}
}

\newcommand{\myproof}[1]{
    \noindent {\bf Proof:} #1 
    
    \vspace{5pt}
}

\newcommand{\myproofof}[2]{
    \noindent {\bf Proof of \cref{#1}:} #2

    \vspace{5pt}
}

\newtheorem{assumptionARpq}{Assumption}

\crefname{assumptionARpq}{Assumption}{Assumptions}
\Crefname{assumptionARpq}{Assumption}{Assumptions}

\crefname{assumptionNVARpq}{Assumption}{Assumptions}
\Crefname{assumptionNVARpq}{Assumption}{Assumptions}

\newcommand{\myassumptionARpq}[3]{%
    \vspace{-5pt}
    \noindent
    \begin{assumptionARpq}[#1]\label{#2}
        #3
    \end{assumptionARpq}
    \vspace{-5pt}
}

\newcommand{\cbr}[1]{\left\{ {#1} \right\}}
\newcommand{\sbr}[1]{\left[ {#1} \right]}
\newcommand{\br}[1]{\left( {#1} \right)}

\newcommand{\myquote}[1]{``{#1}"}

\newcommand{\reals}{\mathbb{R}}
\newcommand{\integers}{\mathbb{Z}}
\newcommand{\naturals}{\mathbb{N}}

\newcommand{\complexnumbers}{\mathbb{C}}

\newcommand{\one}[1]{\mathbf{1}\cbr{ #1 }}

\newcommand{\calL}{\mathcal{L}}

\newcommand{\calI}{\mathcal{I}}

\newcommand{\calH}{\mathcal{H}}

\newcommand{\convdist}{\overset{d}{\rightarrow}}

\newcommand{\calP}{\mathcal{P}}
\newcommand{\wq}{\omega_q}

\newcommand{\ii}{\mathrm{i}}

\newcommand\blfootnote[1]{%
  \begingroup
  \renewcommand\thefootnote{}\footnote{#1}%
  \addtocounter{footnote}{-1}%
  \endgroup
}

\newcommand{\citetwo}[2]{%
  \citet{#1} and \citet{#2}}
\newcommand{\citethree}[3]{%
  \citet{#1}, \citet{#2} and \citet{#3}}

\setlist[itemize]{itemsep=2pt, topsep=3pt, parsep=0pt, partopsep=0pt}
\setlist[enumerate]{itemsep=2pt, topsep=3pt, parsep=0pt, partopsep=0pt}

\newcommand*{\plotPath}{Plots}%

\newcommand{\plotSizeDouble}{0.4}

\newcommand{\authorname}{Marko Mlikota}
\newcommand{\authorinstitution}{Geneva Graduate Institute}

\begin{document}


\title{\vspace*{-0.5 in} \fontsize{18}{20}\selectfont 
	\textbf{Parameter Identification and Inference in Discretely Sampled or Temporally Aggregated Autoregressions}}




\author{\hspace*{1pt}\authorname\blfootnote{
		\setlength{\baselineskip}{4mm} \fontsize{9}{11}\selectfont Correspondence:  Department of International Economics, Geneva Graduate Institute (IHEID), 1202 Geneva, Switzerland. Email: marko.mlikota@graduateinstitute.ch.
		For helpful comments I thank Oriol González-Casasús and Filip Obradović.
		I conducted parts of the analysis in this paper with the help of Anthropic's Claude Fable 5, whose output I thoroughly verified.
		Any errors are my own.} \\[-10pt]
	{\em \small \authorinstitution} }

\date{\vspace*{0pt} {\normalsize This Version: \today} }
\maketitle


\vspace*{-20pt}

\begin{spacing}{1.1}

\begin{abstract}

	\noindent
	I consider an AR($p$) process that is observed every $q$ periods, either as a snapshot (stock variable) or as a sum over the sampling interval (flow variable).
	%
	I first characterize the resulting ARMA process followed by observables.
	Under fairly mild assumptions, 
	I then derive the identified set for general lag lengths $p \in \mathbb{N}$ and sampling frequencies $q \in \mathbb{N}$, 
	I bound its cardinality,
	and I provide an algorithm to compute all candidate points and determine their membership in the identified set.
	%
	My exact but implicit characterization supports the following conjecture that I prove in some settings and verify numerically more broadly:
	(i) the error term-variance is point-identified, 
	(ii) under temporal aggregation, the autoregressive parameters are point-identified, 
	and (iii) under discrete sampling they are point-identified for odd $q$ and identified up to alternating sign for even $q$.
	My analysis supplements existing inference results that show consistency and asymptotic Normality of the Gaussian Maximum Likelihood estimator conditional on point-identification.
	Holding the number of observations fixed, I show that its precision does not necessarily decrease with $q$.
	\todo{REMEMBER: DISABLE TODONOTES \& DELETE ToC BEFORE SHARING}

\end{abstract}



\thispagestyle{empty}







\end{spacing}



\newpage

\clearpage
\setcounter{page}{1}


\section{Introduction}
\label{sec_intro}

Empirical economic analyses must acknowledge the possibility that the frequency referred to by dynamic models may differ from the frequency of the observations available to evaluate these models.
Due to the low frequency of most economic time series data,
the case where the model frequency is higher than the observational frequency is of particular interest.
Yet little is known about the identifiability of parameters in high-frequency models based on low-frequency observations even for ARMA processes, let alone for more complex models;
besides characterizing the aliasing problem introduced by imperfect observation, the long-standing literature on this topic \citep{Telser1967,PalmNijman1984,NijmanPalm1990a} has documented that point-identification fails in some settings and indirectly shown that it holds in a few others.

Building on this literature, I narrow this gap.
I consider a latent scalar process $x_\tau$ evolving as an AR($p$),
\begin{equation} \label{eq_ARp}
    x_\tau = \phi_1 x_{\tau-1} + \dots + \phi_p x_{\tau - p} + e_\tau \; , \quad e_\tau \sim WN(0,v) \; ,
\end{equation}
with $\phi = (\phi_1,\dots,\phi_p)' \in \reals^p$, $p \in \naturals$ and $v \in \reals_{++}$,
and I assume it is observed every $q \in \naturals$ periods, either as a
snapshot (\emph{stock variable}, \cref{eq_ARpq_stock}) or as a sum over the sampling interval
(\emph{flow variable}, \cref{eq_ARpq_flow}):
\begin{align}
    y_t &= x_{tq} \;, \label{eq_ARpq_stock} \\
    y_t &= x_{tq} + x_{tq-1} + \dots + x_{tq-q+1} \;. \label{eq_ARpq_flow}
\end{align}
I dub this observed process AR($p,q$).
I first show that it is equivalent to an ARMA($p,q^*$) with $q^* = \text{floor}(p(q-1)/q)$ in the stock case and an ARMA($p,q^{**}$) with $q^{**} = \text{floor}((p+1)(q-1)/q)$ in the flow case, 
and I characterize its autoregressive polynomial and error-autocovariances in terms of the underlying parameters $(\phi,v)$ of the high-frequency process in \cref{eq_ARp} (\cref{prop_ARpq_stock_observeddynamics,prop_ARpq_flow_observeddynamics}, respectively).
Building on this representation-result, I then derive the identified sets for the true $(\phi,v)$ in both cases
-- denoted by $\calI^{s,1}_{p,q}(\phi,v)$ and $\calI^{f,1}_{p,q}(\phi,v)$, respectively -- 
for general $p$ and $q$ under fairly mild assumptions:
\myassumptionARpq{}{ass_ARpq_ID_stationarity}{
    $x_\tau$ is weakly stationary, i.e. $\phi(z) = \prod_{i=1}^p(1-\lambda_i z)$ with $|\lambda_i| < 1$ for $i=1:p$, or, equivalently, $\phi(z) \neq 0$ for all $|z| \leq 1$.
}

\myassumptionARpq{}{ass_ARpq_ID_actualorderisp}{
    $\phi_p \neq 0$, or, equivalently, $\lambda_i \neq 0$ for $i=1:p$ (since $\phi_p = (-1)^{p+1}\prod_i \lambda_i$).
}

\myassumptionARpq{}{ass_ARpq_ID_distincteigvalqthpowers}{
    $\lambda_i^q \neq \lambda_j^q$ for all $i \neq j$, $i,j = 1:p$.\footnote{
        For $q = 2$, \cref{ass_ARpq_ID_distincteigvalqthpowers} precludes repeated roots and two roots with $\lambda_i = -\lambda_j$.
        For $q = 1$, it reduces to \myquote{no repeated root}.
    }
}

\myassumptionARpq{}{ass_ARpq_ID_normality}{
    $e_\tau \overset{i.i.d.}{\sim} N(0,v)$.
}

My exact but implicit characterizations (\cref{prop_ARpq_stock_ID_generalcase,prop_ARpq_flow_ID_generalcase}, respectively) support the following explicit conclusions: $v$ is point-identified, and $\phi$ is identified at least up to alternating sign, i.e. there is at most one $\tilde\phi$ observationally equivalent to $\phi$ and it equals $\phi^- = (-\phi_1,\phi_2,-\phi_3, ..., (-1)^p\phi_p)'$.
I state these conclusions as \cref{conj_ARpq_stock_ID,conj_ARpq_flow_ID}, distinguishing the stock- and flow-cases.

\myconjecture{AR($p,q$), Stock Variable: Identified Set}{conj_ARpq_stock_ID}{
    Let $x_\tau$ and $y_t$ evolve as in \cref{eq_ARp,eq_ARpq_stock} for some $q \in \naturals$.
    Suppose we only observe $\{y_t\}$ and \cref{ass_ARpq_ID_stationarity,ass_ARpq_ID_actualorderisp,ass_ARpq_ID_distincteigvalqthpowers,ass_ARpq_ID_normality} hold. 
    Then $\calI^{s,1}_{p,q}(\phi,v) = \{(\phi,v)\}$ -- i.e. $(\phi,v)$ is point-identified -- for $q$ odd and $\calI^{s,1}_{p,q}(\phi,v) = \{(\phi,v),(\phi^-,v)\}$ for $q$ even.
}

\myconjecture{AR($p,q$), Flow Variable: Identified Set}{conj_ARpq_flow_ID}{
    Let $x_\tau$ and $y_t$ evolve as in \cref{eq_ARp,eq_ARpq_flow} for some $q \in \naturals$.
    Suppose we only observe $\{y_t\}$ and \cref{ass_ARpq_ID_stationarity,ass_ARpq_ID_actualorderisp,ass_ARpq_ID_distincteigvalqthpowers,ass_ARpq_ID_normality} hold. 
    Then $(\phi,v)$ is point-identified; $\calI^{f,1}_{p,q}(\phi,v) = \{(\phi,v)\}$.
}

Under $q=1$, the distinction between stock and flow is redundant, and we perfectly observe the AR($p$) process in \cref{eq_ARp}, whose parameters are then point-identified even without \cref{ass_ARpq_ID_actualorderisp,ass_ARpq_ID_distincteigvalqthpowers,ass_ARpq_ID_normality}.\footnote{
    I state this well-known result for completeness as \cref{prop_ARpq_ID_q1} See e.g. \citet[Section 8.1]{BrockwellDavis1991} and note my comments in \cref{sec_AR_ID}.
}
More to the point, I prove \cref{conj_ARpq_stock_ID,conj_ARpq_flow_ID} for $q=2$ (\cref{prop_ARp2_stock_ID,prop_ARp2_flow_ID}, respectively).
I also prove \cref{conj_ARpq_stock_ID} when restricting attention to $\phi$ with real roots $\{\lambda_i\}_{i=1}^p$ (\cref{prop_ARpq_stock_ID_realroots}),
and I prove \cref{conj_ARpq_flow_ID} when restricting attention to $\phi$ with real roots and to odd $q$ (\cref{prop_ARpq_flow_ID_realrootsOddq}).
Furthermore, I show that $(\phi^-,v) \in \calI^{s,1}_{p,q}(\phi,v)$ for any even $q$ (\cref{prop_ARpq_stock_ID_qEven_reflectionequiv}), 
while $(\phi^-,\tilde v) \notin \calI^{f,1}_{p,q}(\phi,v)$ for any $\tilde v$ and even $q$ (\cref{prop_ARpq_flow_ID_qEven_reflectionnonequiv}).
Most importantly, I show that both identified sets contain at most $G$ points -- where $G=2^rq^c$ if $q$ is even and $G=q^c$ if $q$ is odd, and $r\in \naturals$ and $c\in\naturals$ are the number of real roots and conjugate pairs of $\phi$, respectively, with $p=r+2c$ --, 
and I provide an algorithm to compute the $G$ candidate points and determine their membership in $\calI^{s,1}_{p,q}(\phi,v)$ and in $\calI^{f,1}_{p,q}(\phi,v)$ for a given $(\phi,v)$ (\cref{prop_ARpq_ID_computationofIDset}).
This enables me to numerically verify that \cref{conj_ARpq_stock_ID,conj_ARpq_flow_ID} hold for many different values of $p$, $q$ and $(\phi,v)$ that satisfy \cref{ass_ARpq_ID_stationarity,ass_ARpq_ID_actualorderisp,ass_ARpq_ID_distincteigvalqthpowers}.

My analysis generalizes earlier results by \citetwo{PalmNijman1984}{NijmanPalm1990a}.
Key to this generalization is my representation of the autoregressive polynomial and error-autocovariances in the observed process in terms of the autoregressive polynomial $\phi$ rather than its roots.

By establishing when point-identification holds, when it fails and under which restriction -- excluding $\phi^-$ -- it can be achieved, my identification analysis supplements existing inference results \citep{PalmNijman1984,NijmanPalm1990b}.
Holding the number of observations fixed, I show that estimation precision does not necessarily decrease with $q$: for instance, under a persistent stock-sampled AR($1,q$) process with $|\phi|=0.9$, it peaks at $q=8$.

Overall, my results suggest that gathering data sampled at a frequency closer to the high frequency at which an underlying model is assumed to evolve does not necessarily improve the identifiability of model parameters nor increase the precision at which they are estimated.
This contrasts with the effects of imperfect observation on forecasting accuracy \citep{ZellnerMontmarquette1971,AmemiyaWu1972,Tiao1972,TiaoWei1976,NijmanPalm1990b} or Granger causality analysis \citep{TiaoWei1976,Weiss1984,Marcellino1999} studied elsewhere in the literature.

\paragraph*{Contribution}

Parameter identification in regularly, but infrequently observed (stock) or temporally aggregated (flow) ARMA processes is a long-standing question in the literature.
\citet{Telser1967} considers an AR($p$) process under both sampling schemes.\footnote{
    My $q$ corresponds to $m+1$ in \citet{Telser1967} and to $m$ in \citet{AmemiyaWu1972,PalmNijman1984,NijmanPalm1990a}.
}
He shows that eigenvalues of the underlying process are pinned down only up to the multi-valued $q$th root, thus motivating my \cref{ass_ARpq_ID_distincteigvalqthpowers}.
He then concludes that point-identification can nevertheless always be achieved based on a statistic that uses the variance of the observed process' residual \citep[p. 494]{Telser1967}, a claim later refuted by \citet[p. 1418]{PalmNijman1984} for the stock case with $p=1$ and $q=2$, which trivially extends to general $p$: 
for any $(\phi,v)$, $y_t = x_{t2}$ with $\phi(L)x_\tau = e_\tau$ is observationally equivalent to 
$y_t = \tilde{x}_{t2}$ with $\phi(-L)\tilde{x}_\tau = e_\tau$ and $\tilde{x}_\tau = (-1)^\tau x_\tau$,
which suggests that we cannot learn the (joint) sign of $(\phi_1,\phi_3,...,\phi_{2\text{ceil}(p/2)-1})'$ in this case.
In contrast, flow sampling eliminates this sign ambiguity \citep[p. 1421]{PalmNijman1984}.
\citet{PalmNijman1984} show that point-identification may fail for more general settings: they consider an ARMA process with exogenous regressors that is observed every $q$ periods as a snapshot (stock) or as a weighted sum over the interval period (flow) with general weights.
%
Disregarding exogenous regressors and restricting attention to AR($p$) processes that are stock-sampled or flow-sampled with equal aggregation weights,
I exactly characterize global identification for any $p$ and $q$.
I thereby generalize the discovery of \citet{PalmNijman1984} that under $q=2$ stock sampling entails a sign ambiguity which flow sampling eliminates, showing that this observation holds for any $p$ and any even $q$ (\cref{prop_ARpq_stock_ID_qEven_reflectionequiv,prop_ARpq_flow_ID_qEven_reflectionnonequiv}).\footnote{
    \citet[p. 1424]{PalmNijman1984} already sketch a general-$q$-analogue of this ambiguity: writing the autoregressive polynomial of the observed process as $\prod_i(1-\tilde\Psi_i L^{q})$, they note that the roots $\alpha_i$ of the underlying autoregressive polynomial solve $\alpha_i^{q}=\tilde\Psi_i$ non-uniquely (in my notation, $\lambda_i^{q}=\mu_i$ with $\mu_i$ the roots of $\rho_q$), but they do not characterize the resulting identified set for $q>2$.
}
My conjectures suggest that despite this counterexample under stock sampling with $q$ even, \citet{Telser1967}'s conclusion that point-identification can be achieved holds in all other cases, against the anticipation of \citet[p.~247]{NijmanPalm1990a} that for large $p$ the number of observationally equivalent models \myquote{can become large}.
For a stock-sampled AR(2), my analysis formalizes the numerical illustration of the identified set for $q=2$ and $q=3$ in \citet{NijmanPalm1985} and their assertion in \citet[p. 244, 246]{NijmanPalm1990a} that $\phi$ and $\phi^-$ are observationally equivalent for any even $q$ and that $\phi$ is point-identified for odd $q$.
%
%
In contrast, my identification results for the flow case are almost entirely new, as it has not been analyzed beyond $p=1$ and $q=2$.

A few studies extend \citet{Telser1967}'s initial characterization of the observed process and provide the foundation for later studies of imperfectly observed ARMA processes.
\citet{AmemiyaWu1972} show that flow-sampled AR($p$) processes follow an ARMA process whose autoregressive polynomial has roots equal to the $q$th powers of the original roots.
\citet{Brewer1973} extends their analysis to a general underlying ARMA process and to both flow- and stock-sampling and derives the order of the implied ARMA process for observables using a counting-argument.
Building on \citetwo{Telser1967}{AmemiyaWu1972}, I explicitly characterize the ARMA process followed by observables in my setting -- a stock- or flow-sampled AR($p$) process --, including the autoregressive polynomial and error-autocovariances.
For their general dynamic regression model, \citet[pp. 1423-1424]{PalmNijman1984}
derive the analogous objects, expressed in terms of the roots of the
underlying autoregressive polynomial. 
Constructions based on coefficients $\phi$ exist as well, but they determine these objects only implicitly, as the solution of a linear system \citep{Brewer1973,Marcellino1999,ForoniEtAl2018}.
My characterization is restricted to the AR($p$) case, but delivers them in closed form as an explicit function of $\phi$. 
This construction is key for my subsequent identification analysis.

Apart from the extensions of the high-frequency process to exogenous regressors and MA-errors discussed above, some studies investigate the observed process under seasonal \citep{Wei1978,Weiss1984} or non-stationary \citep{Weiss1984,HottaVasconcellos1999} high-frequency models.
\citet{JordaMarcellino2004} do so for an ARMA process sampled at a time-varying and potentially stochastic frequency.
Going beyond representation, \citet{DrostNijman1993} show that parameters of a high-frequency GARCH process are point-identified owing to parameters' sign-constraints, a finding qualitatively in line with \cref{conj_ARpq_stock_ID,conj_ARpq_flow_ID}, which suggest that identification in my setting is obtained at least up to sign. 
Parameter identification in imperfectly observed multivariate linear processes is discussed 
in \citetwo{NijmanPalm1990a}{GongEtAl2017} for a VAR(1)
and in \citetwo{Zadrozny2016}{AndersonEtAl2017} for mixed-frequency models (which necessarily feature at least two time series observed at different frequencies).

A related literature studies parameter identification in continuous-time processes.
\citet{Phillips1973} considers a multivariate first-order linear system $Dx(t) = Ax(t) + \xi(t)$ 
-- where $D$ is the time-derivative operator, $A$ is a square matrix and $\xi(t)$ is a noise process --
that is observed at discrete time points with inter-observational spacing $h \in \reals_{++}$.\footnote{
    This underlying system corresponds to the Ornstein Uhlenbeck process $dx(t) = A x(t) dt + \xi (dt)$.
}
The observations $x(th)$ then follow a VAR(1) with autoregressive matrix $B=e^{Ah}$, whose inversion yields countably infinitely many parameter-matrices $A$.
\citet{Phillips1973} concludes that there are countably infinitely many observationally equivalent models and discusses sufficient restrictions for identification.
\citet{HansenSargent1983} show that the number of observationally equivalent models is finite since not every $A$ can support a positive semidefinite error-term covariance $Cov\br{ \int_0^1 \xi(s) ds}$.\footnote{
    They consider a more general underlying process that nests the one in \citet{Phillips1973}.
}
In my discrete-time setting, the analogue of the mapping $A \to e^{Ah}$ is the mapping $\lambda_i \to \lambda_i^q$ from roots of $\phi$ to roots of the observed process' autoregressive polynomial, whose inversion yields finitely many solutions and bounds the cardinality of the identified set.\footnote{
    \citet{NijmanPalm1990a} draw a similar analogy between the continuous-time setting of \citetwo{Phillips1973}{HansenSargent1983} and their imperfectly observed VAR(1).
}
This bound is affected by the distinction between sampling at even or odd frequencies $q$, which is without analogue in the continuous-time environment.
Akin to the insight of \citet{HansenSargent1983}, it is the requirement to support a positive error-variance that further trims the identified set.
%
In my setting this requirement seems to reduce the identified set all the way to yield point-identification or identification up to alternating sign.

\citethree{Telser1967}{PalmNijman1984}{NijmanPalm1985} outline possible approaches to estimate $(\phi,v)$.
%
\citet[p. 492]{Telser1967} proposes a method-of-moments estimator for $\phi$.\footnote{
    He notes, however, that inverting the observed process' autoregressive polynomial is a multi-valued problem and suggests to resolve it by means of an additional, ad-hoc residual-variance ratio (see his Eq. (31); the residual representations entering it are generalized to arbitrary $p$ and $q$ in Eqs. (30$'$) and (33)), which he cautions is hard to estimate in practice even in the simplest case (see footnote 5 on p. 493).
}
\citet{PalmNijman1984} outline direct Maximum Likelihood (ML) estimation of the restricted ARMA process, circumventing the need to evaluate the likelihood via the Kalman filter as suggested in \citet{NijmanPalm1985}.
My analytical characterization of the mapping between $(\phi,v)$ and the autoregressive polynomials and error-autocovariances in the observed process makes both routes operational without root-finding,
whereby my identification analysis clarifies the roles of the autoregressive polynomial and the error-autocovariances in the implied moment conditions.\footnote{
    The autoregressive block, $\rho_q=\rho_q(\phi)$, is free of $v$ and depends on the roots of $\phi$ only through their $q$th powers.
    By \cref{lemma_aliasingbound}, it therefore determines $\phi$ only up to $G$ candidate parameter vectors. 
    The $q^*+1$ (stock) or $q^{**}+1$ (flow) error-autocovariances contain all remaining second-moment information; the implied moments correspond to the checks in \cref{prop_ARpq_ID_computationofIDset} that I use to numerically verify \cref{conj_ARpq_stock_ID,conj_ARpq_flow_ID}.
    %
}  
Moreover, my analysis clarifies when we obtain point-identification, conditional on which \citetwo{PalmNijman1984}{NijmanPalm1990b} derive the asymptotic distribution of the Gaussian ML estimator of $(\phi,v)$.

A related literature discusses estimation efficiency under imperfect observation \citep{PalmNijman1984,NijmanPalm1990b,PierseSnell1995,PonsSanso2005}.
Holding the number of high-frequency periods fixed, \citetwo{PalmNijman1984}{NijmanPalm1990b} establish that coarser sampling necessarily loses information.
Holding instead the number of observations fixed, as is relevant when a given sample is modeled at alternative underlying frequencies, I show that coarser sampling can be more informative about the parameters.
Specifically, for a stock-sampled AR($1,q$) process, I derive the sampling interval $q^*$ that maximizes information about $\phi$ in closed form.

\paragraph*{Outline} 

The rest of this paper is structured as follows. 
I derive the observed process in \cref{sec_observeddynamics}. 
In turn, \cref{sec_AR_ID} treats identification and \cref{sec_est} discusses estimation.
These sections are accompanied by \cref{appsec_observeddynamics,appsec_ID,appsec_est}, respectively.
\cref{sec_conclusion} concludes.
Propositions and Lemmas are numbered by the order of their proofs in the Appendix.




\section{Representation}
\label{sec_observeddynamics}




In this section, I characterize the observed process.
The resulting \cref{prop_ARpq_stock_observeddynamics,prop_ARpq_flow_observeddynamics}, the supporting \cref{lemma_PsiqRewriting} and the surrounding definitions provide the foundation for my subsequent identification analysis in \cref{sec_AR_ID}.
Details are in \cref{appsec_observeddynamics}.

Let $\phi_0 = -1$ and write $\phi(z) = -\sum_{j=0}^{p}\phi_j z^j = \prod_{i=1}^p (1-\lambda_i z)$.
For some $q \in \naturals$, let $\wq = e^{2\pi \ii/q}$.
Note that $\wq^q = 1$.
%
%
Define
$$ 
    \Psi_q(z) = \prod_{j=0}^{q-1} \phi(\wq^j z) \; .
$$
\cref{lemma_PsiqRewriting} shows that $\Psi_q(z)$ only contains powers of $z$ that are divisible by
$q$,
that there exists therefore a polynomial $\rho_q$ with degree $\leq p$ s.t. $\rho_q(z^q) = \Psi_q(z)$,
and that $\rho_q$ has real coefficients.
Given this, define 
$$
    c_q(z) = \prod_{j=1}^{q-1} \phi(\wq^j z) \; .
$$
Given this definition, $c_q(z)$ is a polynomial of degree $p(q-1)$ with coefficients $c_{q,0}=1,c_{q,1},\dots,c_{q,p(q-1)}$.
Since $c_q(z) = \rho_q(z^q) / \phi(z)$ and both $\rho_q$ and $\phi$ are real, $c_q(z)$ is real as well.

\mylemma{}{lemma_PsiqRewriting}{
    Let $\phi$ be a real polynomial with roots $\lambda_1,\dots,\lambda_p$,
    and let $q\in \naturals$.
    Then $\exists$ a polynomial 
    $\rho_q(z) = \prod_{i=1}^{p} (1-\lambda_i^q z) = 1 - \sum_{k=1}^{p} \rho_{q,k} z^k$
    with real coefficients 
    s.t. 
    $\Psi_q(z) = \rho_q(z^q)$.
    For $q=1$, $\rho_1 = \phi$.
}

\myproposition{Stock Case: Observed Dynamics}{prop_ARpq_stock_observeddynamics}{
    Let $x_\tau$ and $y_t$ evolve as in \cref{eq_ARp,eq_ARpq_stock} for some $q\in \naturals$, with $e_\tau \sim WN(0,v)$. 
    Let $\{\lambda_i\}_{i=1}^p$ be the roots of $\phi(z) = \prod_{i=1}^p (1-\lambda_i z)$.
    Then $y_t$ follows an ARMA($p,q^*$) with $q^* = \text{floor}\br{ p(q-1)/q  }$:
    $\rho_q(L) y_t = u_t$,
    where $u_t$ has mean zero and autocovariances
    $\gamma_u(h) = v\sum_{k=0}^{p(q-1)-qh} c_{q,k} c_{q,k+qh}$ for $0 \leq h \leq q^*$ and zero for $h > q^*$.
}

\myproof{
    Multiplying $\phi(L)x_\tau = e_\tau$ by $c_q(L)$ yields $\Psi_q(L)x_\tau = c_q(L)e_\tau$. 
    Since $\Psi_q(L) = \rho_q(L^q)$ by \cref{lemma_PsiqRewriting} and
    $L^{qk}x_{tq} = x_{(t-k)q} = y_{t-k}$, 
    evaluating at $\tau = tq$ gives $\rho_q(L)y_t = u_t$ with $u_t = c_q(L)e_{tq} = \sum_{k=0}^{p(q-1)}c_k e_{tq-k}$.
    Note that $u_t = \sum_k c_k e_{tq-k}$ and $u_{t-h} = \sum_{k'} c_{k'} e_{tq-qh-k'}$ share shock $e_{tq-k}$ iff $k' = k-qh$. 
    Hence, $\gamma_u(h) = v \sum_{k} c_k c_{k-qh} = v\sum_{k=0}^{p(q-1)-qh} c_{k+qh}c_k$, which is zero as soon as $qh > p(q-1)$, i.e. for $h > q^*$.
    $\blacksquare$
}

For the flow case, define also
$$ s_q(z) = 1 + z + \dots + z^{q-1} \quad \text{and} \quad S_q(z) = s_q(z)s_q(z^{-1}) \; , $$
and let $X_\tau = s_q(L)x_\tau$ s.t. $y_t = X_{tq}$. 
Note that $s_q(1)=q$ and $s_q(z) = (1-z^q)/(1-z)$ for $z \neq 1$, so the zeros of
$s_q$ are exactly the $q$th roots of unity other than $1$,
while $s_q(\lambda) \neq 0$ and $s_q(1/\lambda) \neq 0$ for $0<|\lambda|<1$. 
Also, note that $\gamma_X(h)=\sum_{|m|\leq q-1}(q-|m|)\gamma_x(h+m)$,\footnote{
    Since $X_\tau = s_q(L)x_\tau = \sum_{l=0}^{q-1} x_{\tau-l}$, we have $\gamma_X(h) = \sum_{l=0}^{q-1}\sum_{k=0}^{q-1} \gamma_x(h+k-l)$, and there are $q-|m|$ pairs with $k-l=m$.
} 
and by similar reasoning we can see that $S_q(z) = \sum_{|m| \leq q-1}(q-|m|)z^m$. 
Define also the real polynomial $d_q(z) = c_q(z)s_q(z)$ with $\deg d = p(q-1) + q - 1 = (p+1)(q-1)$ and coefficients $\{d_{q,k}\}$.
If the flow is observed as an average 
-- $y_t = (x_{tq}+\dots+x_{tq-q+1})/q$ --,
then $S_q(z)/q^2$ replaces $S_q(z)$ throughout and all results hold unchanged.

\myproposition{Flow Case: Observed Dynamics}{prop_ARpq_flow_observeddynamics}{
    Let $x_\tau$ and $y_t$ evolve as in \cref{eq_ARp,eq_ARpq_flow} for some $q\in \naturals$, with $e_\tau \sim WN(0,v)$.
    Let $\{\lambda_i\}_{i=1}^p$ be the roots of $\phi(z) = \prod_{i=1}^p (1-\lambda_i z)$.
    Then $y_t$ follows an ARMA($p,q^{**}$) with $q^{**} = \text{floor}\br{ (p+1)(q-1)/q }$:
    $\rho_q(L) y_t = u_t$,
    where $u_t$ has mean zero and autocovariances
    $\gamma_u(h) = v\sum_{k=0}^{(p+1)(q-1)-qh} d_{q,k} d_{q,k+qh}$ for $0 \leq h \leq q^{**}$ and zero for $h > q^{**}$.
}

\myproof{
    Applying $s_q(L)$ to $\phi(L)x_\tau = e_\tau$ gives $\phi(L)X_\tau = s_q(L)e_\tau$.
    Multiplying then by $c_q(L)$ yields $\Psi_q(L)X_\tau = d_q(L)e_\tau$.
    Since $\Psi_q(L) = \rho_q(L^q)$ by \cref{lemma_PsiqRewriting} and $L^{qk}X_{tq} = X_{(t-k)q} = y_{t-k}$, evaluating at $\tau = tq$ gives $\rho_q(L)y_t = u_t$ with
    $u_t = d_q(L)e_{tq} = \sum_{k=0}^{(p+1)(q-1)}d_k e_{tq-k}$.
    The autocovariance computation is then as in \cref{prop_ARpq_stock_observeddynamics}, with $\{d_k\}$ in place of $\{c_k\}$, and $\gamma_u(h) = 0$ iff $qh > (p+1)(q-1)$.
    $\blacksquare$
}

\citet{PalmNijman1984} already characterize the autoregressive polynomial and the error-autocovariances in the observed process for general $p$ and $q$ and for a more general underlying ARMA process with exogenous regressors
based on a similar annihilator-polynomial technique introduced in \citetwo{Telser1967}{AmemiyaWu1972}.\footnote{
    \citet{AmemiyaWu1972} also characterize the autoregressive polynomial for the flow case.
}
Their construction is root-based: the annihilator polynomial is assembled from the roots of $\phi(z)$.
\citet{Brewer1973}, \citet{Marcellino1999} and \citet{JordaMarcellino2004} work with the coefficients of $\phi(z)$ instead, but they determine the annihilator polynomial only implicitly, as the solution of a linear system.
My definition of $\Psi_q(z)$ and the resulting $\rho_q(z)$, $c_q(z)$ and $d_q(z)$ deliver the autoregressive polynomial and the residual autocovariances in closed form in the coefficient-vector $\phi$, requiring neither the roots of $\phi(z)$ nor the solution of a linear system.
This distinction is key to my extension of the identification analysis of \citethree{PalmNijman1984}{NijmanPalm1985}{NijmanPalm1990a} in \cref{sec_AR_ID}.

\todo{see here commented-out two paragraphs that go into more detail on both (i) root-based methods, (ii) the implicite nature of existing coefficient-based methods}

\paragraph*{Properties of the Observed Process}

For a stationary process $a_t$ with autocovariances $\gamma_a(h)$, define the Autocovariance Generating Function (ACGF) $\Gamma_a(z) = \sum_{h=-\infty}^{\infty} \gamma_a(h) z^h$ for $z \neq 0$.
For $e_\tau \sim WN(0,v)$, we have $\Gamma_e(z) = v$.
Given $\Gamma_a(z)$, for $b_t = B(L)a_t$, we have $\Gamma_b(z) = B(z)B(z^{-1})\Gamma_a(z)$ \citep[Section 3.5, Eq. (3.5.4) - (3.5.5)]{BrockwellDavis1991},
and, therefore, we have $\Gamma_x(z)= v/\Phi(z)$ for $x_\tau$ and any $(\phi,v) \in \Theta_p$, where for $z \neq 0$ we define the symmetric Laurent polynomial $\Phi(z) = \phi(z)\phi(z^{-1}) = \sum_{j,k}\phi_j\phi_k z^{j-k}$.

Three properties of the observed ARMA process matter for estimation in \cref{sec_est}.
First, it is stationary under \cref{ass_ARpq_ID_stationarity}: the roots of $\rho_q(z)$ are $\lambda_i^{-q}$, which lie outside the unit circle iff $|\lambda_i|<1$.
Second, it is Gaussian under \cref{ass_ARpq_ID_stationarity,ass_ARpq_ID_normality}: then $x_\tau$ is a mean-square convergent linear combination of $\{e_{\tau-j}\}_{j\geq0}$, so $y_t$ is a Gaussian process, and, provided its ARMA representation is invertible, the innovations are linear combinations of current and past $y_t$ and therefore Gaussian white noise, i.e. i.i.d.
Third, it is invertible under \cref{ass_ARpq_ID_stationarity}.
This is proven for the flow case by \citet[p.~629]{AmemiyaWu1972}.
Their argument extends directly to the stock case.\footnote{
    In my notation, 
    since $\gamma_u(h)=0$ for $h>q^*$, $u_t$ has a moving-average representation $u_t=\theta_q(L)\varepsilon_t$ of order at most $q^*$ with $\varepsilon_t\sim WN(0,\sigma^2)$ \citep[Proposition 3.2.1]{BrockwellDavis1991}, and $\theta_q$ can be chosen with all roots outside the unit circle as long as the autocovariance generating function $\Gamma_u(w)=\sum_h\gamma_u(h)w^h=\sigma^2\theta_q(w)\theta_q(w^{-1})$ has no zeros on the unit circle \citep[Proposition 4.4.2]{BrockwellDavis1991}.
    This is the case for a stock variable: by the proof of \cref{prop_ARpq_stock_observeddynamics}, $\gamma_u(h)/v$ is the coefficient of $z^{qh}$ in $c_q(z)c_q(z^{-1})$, and since $\frac{1}{q}\sum_{j=0}^{q-1}\wq^{jm}$ equals one if $q$ divides $m$ and zero otherwise, we have, for $|z|=1$,
    $$ \Gamma_u(z^q) = \frac{v}{q}\sum_{j=0}^{q-1} c_q(\wq^j z)\,c_q(\wq^{-j}z^{-1}) = \frac{v}{q}\sum_{j=0}^{q-1} \big|c_q(\wq^j z)\big|^2 > 0 \; , $$
    where the second equality uses that $c_q$ is real and $z^{-1}=\bar z$, and the inequality that the zeros of $c_q$, $\wq^{-j}/\lambda_i$, lie outside the unit circle under \cref{ass_ARpq_ID_stationarity}.
}


\section{Identification}
\label{sec_AR_ID}

Building on \cref{sec_observeddynamics}, I now turn to identification.
\cref{appsec_ID} proves the propositions from this section.
Define the parameter space
$$ \Theta_p = \Big\{ (\phi,v)' : \; \phi(z) \neq 0 \; \text{for all} \; |z| \leq 1\;, \quad v > 0 \Big\} \; . $$
Two points of $\Theta_p$ are \emph{observationally equivalent} if they induce the same distribution of $\{y_t\}$. 
The \emph{identified set} of a true value $(\phi,v)$, is the set of all points of $\Theta_p$ observationally equivalent to it,
and $(\phi,v)$ is \emph{point-identified} if its identified set is the singleton $\{(\phi,v)\}$. 
I write $\calI^{s,1}_{p,q}(\phi,v)$ for the identified set of $(\phi,v)$ if $y_t$ is a stock variable and $\calI^{f,1}_{p,q}(\phi,v)$ if $y_t$ is a flow variable.
Note that if $y_t$ is Gaussian with mean zero, observational equivalence corresponds to equality of the autocovariance functions $\gamma_y(h)$.
Also, note that writing $(\tilde \phi, \tilde v) \in \Theta_p$ implies that $\tilde\phi$ satisfies \cref{ass_ARpq_ID_stationarity}.

It is well-known that $(\phi,v)$ is point-identified under $q=1$, in which case we observe the underlying AR($p$) process in every period.
I state this result as \cref{prop_ARpq_ID_q1}.\footnote{
    This result appears implicitly in \citet[Section 8.1]{BrockwellDavis1991} and underlies their Yule-Walker equations that characterize the autocovariance function of the AR($p$) process.
    My proof of \cref{prop_ARpq_ID_q1} proceeds via the ACGF rather than just the first $p+1$ autocovariances, in line with my extensions to the environments with infrequent observation ($q>1$).
}

\myrepproposition{Identification of AR($p,1$)}{prop_ARpq_ID_q1}{
    Let $x_\tau$ evolve as in \cref{eq_ARp} with $e_\tau \sim WN(0,v)$.
    Suppose we observe $y_t = x_t$ and \cref{ass_ARpq_ID_stationarity} holds. 
    Then $(\phi,v)$ is point-identified; $\calI^{s,1}_{p,1}(\phi,v) = \calI^{f,1}_{p,1}(\phi,v) = \{(\phi,v)\}$.
}

To investigate identification under $q>1$ in \cref{subsec_AR_ID_stock,subsec_AR_ID_flow,subsec_AR_ID_conjectures}, 
I define a few more objects below, 
and I state some supporting claims in \cref{appsubsec_ID_lemmas}.
Let $\phi^-(z) = \phi(-z)$, with coefficients $\phi_j^- = (-1)^j \phi_j$ and roots $\lambda_i^- = -\lambda_i$.
Note that $\phi^-$ is stationary iff $\phi$ is. 
For a Laurent polynomial $Q$, define 
$[Q]_q(z) = \frac{1}{q}\sum_{j=0}^{q-1} Q(\wq^j z)$,
which keeps only the powers of $z$ divisible by $q$ (by the same argument as for $\Psi_q$ in the proof of \cref{lemma_PsiqRewriting}).

My identification analysis relies heavily on \cref{lemma_aliasingbound}.
Whereas other supporting lemmas are in \cref{appsubsec_ID_lemmas}, I state \cref{lemma_aliasingbound} below and delegate only its proof to \cref{appsubsec_ID_lemmas}.

\myreplemma{}{lemma_aliasingbound}{
    Let $\phi$ and $\tilde\phi$ be real polynomials with degrees $\leq p$ and constant term $1$.
    Let $q \in \naturals$.
    Suppose $\phi$ satisfies \cref{ass_ARpq_ID_stationarity,ass_ARpq_ID_actualorderisp,ass_ARpq_ID_distincteigvalqthpowers}
    and the roots of $\tilde\phi$ can be re-indexed s.t. $\tilde\lambda_i^q = \lambda_i^q$, $i=1:p$.
    Then 
    \begin{itemize}
        \item[(i)] $\tilde\phi$ also satisfies \cref{ass_ARpq_ID_stationarity,ass_ARpq_ID_actualorderisp,ass_ARpq_ID_distincteigvalqthpowers};
        \item[(ii)] the re-indexing is unique, and $\tilde\lambda_i = \wq^{k_i}\lambda_i$ for a unique $k_i \in \{0,\dots,q-1\}$;
        \item[(iii)] if $\lambda_{i'} = \overline{\lambda_i}$ with $i' \neq i$, then
        $\tilde\lambda_{i'} = \overline{\tilde\lambda_i}$;
        \item[(iv)] if $\lambda_i \in \reals$, then $\tilde\lambda_i = \lambda_i$ if $q$ is odd and $\tilde\lambda_i \in \{\lambda_i, -\lambda_i\}$ if $q$ is even; 
        \item[(v)] if $\phi$ has $r$ real roots and $c$ conjugate pairs ($r+2c = p$), the number of root multisets $\{\tilde\lambda_i\}_{i=1}^p$ consistent with (iii) and (iv) --- or, equivalently, the number of $\tilde\phi$ consistent with $\tilde\lambda_i^q = \lambda_i^q$ --- is exactly $2^r q^c$ for $q$ even and $q^c$ for $q$ odd.
    \end{itemize}
}

The count in statement (v) is a property of the roots alone and does not depend on the observation scheme: under \cref{ass_ARpq_ID_distincteigvalqthpowers} it equals the number of real matrices $\tilde\Pi$ with $\tilde\Pi^q=\Pi^q$ for any real matrix $\Pi$ with eigenvalues $\{\lambda_i\}_{i=1}^p$, such as the companion form-matrix of $\phi$.
For some cases, it boils down to existing counts in the literature.
\citet{NijmanPalm1990a} cover the cases in which all roots are real, all roots are complex, or exactly one root is real -- $(r,c)=(p,0)$, $(0,p/2)$ for even $p$ and $(1,(p-1)/2)$ for odd $p$ -- as well as both cases $(r,c)\in\{(2,0),(0,1)\}$ for $p=2$ and any $q$, which \citet{NijmanPalm1985} illustrate numerically for $q=2$ and $q=3$.\footnote{
    For a $K$-dimensional VAR(1) that is observed every $m$th period, \citet[p.~241]{NijmanPalm1990a} note that the candidate matrices $\tilde\Pi$ number $2^K$ if all eigenvalues are real and $m$ is even, one if they are real and $m$ is odd, and at most $m^{K/2}$ ($K$ even) or $2m^{(K-1)/2}$ ($K$ odd) if they are complex; with $K=p$, these are the counts stated in the main text, the complex ones as upper bounds that coincide with $G$ except for one real root and odd $q$.
    For the univariate ARMA(2,1) and AR(2), their Table~II (p.~245) lists one candidate for real roots and odd $m$, four for real roots and even $m$, and $m$ for complex roots; \citet[pp.~151--152]{NijmanPalm1985} enumerate the four sign choices for real roots under $m=2$ and the two and three candidates for complex roots under $m=2$ and $m=3$, and plot the expected log-likelihood over them.
}
\cref{lemma_aliasingbound} covers the mixed case of $r$ real roots and $c$ conjugate pairs for any $p$ and, through statements (ii)--(iv), delivers the candidates themselves.
Whether a candidate is observationally equivalent is a separate question, for which the observation scheme matters. 
\citet[p.~243]{NijmanPalm1990a} note that their VAR(1) results do not carry over to univariate models with higher lags because only one coordinate of the state vector in the companion form-VAR(1) is observed.
Building on \cref{lemma_aliasingbound} and in particular its statements (ii)--(iv), I derive the set of observationally equivalent candidates in \cref{subsec_AR_ID_stock,subsec_AR_ID_flow}.




\subsection{Regular but Infrequent Observation: the Stock Case}
\label{subsec_AR_ID_stock}

\myrepproposition{Stock Case: Identification of
Observed Dynamics}{prop_ARpq_stock_observeddynamics_ID}{
    Let $x_\tau$ and $y_t$ evolve as in \cref{eq_ARp,eq_ARpq_stock} for some $q\in \naturals$, with $e_\tau \sim WN(0,v)$.
    Suppose we only observe $\{y_t\}$ and \cref{ass_ARpq_ID_stationarity,ass_ARpq_ID_actualorderisp,ass_ARpq_ID_distincteigvalqthpowers} hold.
    Consider a $(\tilde\phi,\tilde v) \in
    \Theta_p$ that is observationally equivalent to $(\phi,v)$. 
    Then
    \begin{itemize}
        \item[(i)] the roots of $\tilde\phi$ satisfy $\{\tilde\lambda_i^q\}_{i=1}^p = \{\lambda_i^q\}_{i=1}^p$ and can be uniquely re-indexed s.t. $\tilde\lambda_i^q = \lambda_i^q$, $i=1:p$;
        \item[(ii)] with this re-indexing, $K_i(\tilde\phi,\tilde v) =
        K_i(\phi,v)$, $i=1:p$;
        \item[(iii)] $\tilde\rho_q = \rho_q$ and $\Gamma_{\tilde u} =
        \Gamma_u$.
    \end{itemize}
}

\myrepproposition{Stock Case: Identified Set}{prop_ARpq_stock_ID}{
    Let $x_\tau$ and $y_t$ evolve as in \cref{eq_ARp,eq_ARpq_stock} for some $q \in \naturals$.
    Suppose we only observe $\{y_t\}$ and \cref{ass_ARpq_ID_stationarity,ass_ARpq_ID_actualorderisp,ass_ARpq_ID_distincteigvalqthpowers,ass_ARpq_ID_normality} hold. 
    Then 
    \begin{itemize}
        \item[(i)] $\calI^{s,1}_{p,q}(\phi,v) = \Big\{ (\tilde\phi,\tilde v) \in \Theta_p  : \; \tilde\lambda_i^q = \lambda_i^q \; , \; K_i(\tilde\phi,\tilde v) = K_i(\phi,v) \; , \; i=1:p \; \text{(re-indexed)} \Big\}$;
        \item[(ii)] if $q$ is even, then $(\phi^-,v)$ is observationally equivalent to $(\phi,v)$;\footnote{
            This holds even without \cref{ass_ARpq_ID_actualorderisp,ass_ARpq_ID_distincteigvalqthpowers}.
        }
        \item[(iii)] if $q=2$, then $\calI^{s,1}_{p,2}(\phi,v)=\cbr{ (\phi,v),(\phi^-,v) }$, and $\phi \neq \phi^{-}$;
        \item[(iv)] if all roots $\lambda_i$ are real, then $\calI^{s,1}_{p,q}(\phi,v) =\{(\phi,v)\}$ for $q$ odd and $\calI^{s,1}_{p,q}(\phi,v) = \cbr{ (\phi,v),(\phi^-,v) }$ for $q$ even.
    \end{itemize}
}

\citet{PalmNijman1984} already establish that under $q=2$ stock sampling entails a sign ambiguity, and \citet[p.~406]{NijmanPalm1990b} and \citet[p.~891]{JordaMarcellino2004} note that for an AR(1) it does so under any even $q$.
My \cref{prop_ARpq_stock_ID_qEven_reflectionequiv} shows that this holds for any $p$ and any even $q$.
\citet{PalmNijman1984} also show that flow sampling eliminates this sign ambiguity under $q=2$.
Further below, in \cref{prop_ARpq_flow_ID_qEven_reflectionnonequiv}, I generalize this to any even $q$.

For a stock-sampled AR(2), \citet[p. 245]{NijmanPalm1990a} already assert that exactly two observationally equivalent parameter vectors that differ by a joint sign flip of the roots arise under any even $q$, and they note that the second autoregressive coefficient therefore remains globally identified -- in line with my characterization.
Their argument establishes that the sign-flipped candidate is observationally equivalent, but the elimination of all other candidate root choices rests on the information content of the observed moving-average parameter and is verified only in examples.\footnote{
    For the ARMA(2,1) case, they note that they \myquote{cannot exclude the possibility} that further solutions are compatible with the observed moments (p. 244).
}
Moreover, they do not state whether the error-term variance is common across the two equivalent models. 
My \cref{prop_ARp2_stock_ID} turns their assertion into an exact characterization for $q=2$: it establishes that the two points exhaust the identified set, it shows that the error-term variance is point-identified, and it allows for general $p$.
In \cref{prop_ARpq_stock_ID_realroots}, restricting attention to real roots, I extend the exact two-point characterization to general $p$ and any even $q$, and I prove point-identification for any odd $q$. 
For $p=2$ and real, unequal roots, \citet[p.~244]{NijmanPalm1990a} already argue the odd-$q$ part: the $q$th powers of the roots are identified, for odd $q$ the only real polynomial of degree two whose roots have these $q$th powers is $\phi$ itself, and the error-term variance then follows from the observed one.
In \citet[pp. 151-152]{NijmanPalm1985}, they already illustrate both parts of this result numerically for an AR(2): under $q=2$, the two sign-flipped root pairs $(.7,.5)$ and $(-.7,-.5)$ attain the same likelihood maximum while the mixed-sign pairs do not, and under $q=3$ the true parameter seems to be the unique global maximizer of the likelihood.

\subsection{Temporal Aggregation: the Flow Case}
\label{subsec_AR_ID_flow}

\myrepproposition{Flow Case: Identification of
Observed Dynamics}{prop_ARpq_flow_observeddynamics_ID}{
    Let $x_\tau$ and $y_t$ evolve as in \cref{eq_ARp,eq_ARpq_flow} for some $q\in \naturals$, with $e_\tau \sim WN(0,v)$.
    Suppose we only observe $\{y_t\}$ and \cref{ass_ARpq_ID_stationarity,ass_ARpq_ID_actualorderisp,ass_ARpq_ID_distincteigvalqthpowers} hold.
    Consider a $(\tilde\phi,\tilde v) \in
    \Theta_p$ that is observationally equivalent to $(\phi,v)$.
    Then
    \begin{itemize}
        \item[(i)] the roots of $\tilde\phi$ satisfy $\{\tilde\lambda_i^q\}_{i=1}^p = \{\lambda_i^q\}_{i=1}^p$ and can be uniquely re-indexed s.t. $\tilde\lambda_i^q = \lambda_i^q$, $i=1:p$;
        \item[(ii)] with this re-indexing, $K_i^X(\tilde\phi,\tilde v) = K_i^X(\phi,v)$, $i=1:p$, where $K_i^X(\tilde\phi,\tilde v) = K_i(\tilde\phi,\tilde v)S_q(\tilde\lambda_i)$;
        \item[(iii)] $\tilde\rho_q = \rho_q$ and $\Gamma_{\tilde u} = \Gamma_u$.
    \end{itemize} 
    Moreover, $\gamma_y(h) = \sum_{i=1}^p K_i^X\lambda_i^{qh}$ for $h\geq 1$, where $K_i^X \neq 0$.
}

\myrepproposition{Flow Case: Identified Set}{prop_ARpq_flow_ID}{
    Let $x_\tau$ and $y_t$ evolve as in \cref{eq_ARp,eq_ARpq_flow} for some $q \in \naturals$.
    Suppose we only observe $\{y_t\}$ and \cref{ass_ARpq_ID_stationarity,ass_ARpq_ID_actualorderisp,ass_ARpq_ID_distincteigvalqthpowers,ass_ARpq_ID_normality} hold. 
    Then 
    \begin{itemize}
        \item[(i)] $\calI^{f,1}_{p,q}(\phi,v) = \Big\{ (\tilde\phi,\tilde v) \in \Theta_p  : \; \text{} \; \tilde\lambda_i^q = \lambda_i^q \; , \; K_i^X(\tilde\phi,\tilde v) = K_i^X(\phi,v) \; \; , \; i=1:p \; \text{(re-indexed)} \; ,  \; \gamma_{\tilde y}(0) = \gamma_y(0) \Big\}$, where $\gamma_y(0) = \sum_{i=1}^p K_i(\phi,v)\sbr{ q + 2\sum_{m=1}^{q-1}(q-m)\lambda_i^m }$ and likewise with tildes;
        \item[(ii)] if $q=2$, then $\calI^{f,1}_{p,2}(\phi,v) = \{(\phi,v)\}$;
        \item[(iii)] if $q$ is even, then $(\phi^-,\tilde v)$ is not observationally equivalent to $(\phi,v)$ for any $\tilde v$;\footnote{
            This holds even without \cref{ass_ARpq_ID_normality}.
        }
        \item[(iv)] if all roots $\lambda_i$ are real and $q$ is odd, then $\calI^{f,1}_{p,q}(\phi,v) = \{(\phi,v)\}$.
    \end{itemize}
}

In the flow case with $q=2$, $S_2(1) = 4 \neq 0 = S_2(-1)$ breaks the $z \to -z$ symmetry of the observed process obtained under stock-sampling.
This is recognized by \citet{PalmNijman1984} and underpins my exact identification result for that case in \cref{prop_ARp2_flow_ID}.
\cref{prop_ARpq_flow_ID_qEven_reflectionnonequiv} shows that this holds more generally for even $q$: $S_q(1) = q^2 \neq 0 = S_q(-1)$.

The flow case has otherwise received little attention: \citet[p.~1421]{PalmNijman1984} note the consequence for $p=1$ and $q=2$ in a single sentence (\myquote{the sign of $\rho$ is determined here}) without derivation, \citet[p.~407]{NijmanPalm1990b} state it for the flow-sampled AR(1) under $q=2$, \citet[pp.~150--151]{NijmanPalm1985} illustrate it for a flow-sampled ARMA(1,1) and MA(1) under $q=2$ by plots of the expected likelihood, and \citet[p.~247]{NijmanPalm1990a} leave \myquote{observations on flow variables} as a \myquote{straightforward} extension.
\cref{prop_ARpq_flow_observeddynamics_ID,prop_ARpq_flow_ID} and \cref{conj_ARpq_flow_ID} are the first results for $p>1$ or $q>2$.




\subsection{Computation of Identified Sets \& Conjectures}
\label{subsec_AR_ID_conjectures}

\cref{prop_ARpq_stock_ID_generalcase,prop_ARpq_flow_ID_generalcase} characterize the identified sets for general lag lengths $p$ and sampling frequencies $q$.
They show that any observationally equivalent $(\tilde\phi,\tilde v) \in \Theta_p$ has roots $\tilde\lambda_i^q = \lambda_i^q$.
\cref{lemma_aliasingbound} bounds the number of candidates $\tilde\phi$ that satisfy this root-equality and provides a recipe to compute all such candidates.
Taken together, these three results provide an upper bound for the identified sets and an algorithm to compute them for different values of $p$, $q$ and $(\phi,v)$ that satisfy \cref{ass_ARpq_ID_stationarity,ass_ARpq_ID_actualorderisp,ass_ARpq_ID_distincteigvalqthpowers}: see \cref{prop_ARpq_ID_computationofIDset}.

\myrepproposition{Upper Bound \& Computation of Identified Set}{prop_ARpq_ID_computationofIDset}{
    Let $x_\tau$ and $y_t$ evolve as in \cref{eq_ARp,eq_ARpq_stock} (stock) or as in \cref{eq_ARp,eq_ARpq_flow} (flow) for some $q \in \naturals$.
    Suppose we only observe $\{y_t\}$ and \cref{ass_ARpq_ID_stationarity,ass_ARpq_ID_actualorderisp,ass_ARpq_ID_distincteigvalqthpowers,ass_ARpq_ID_normality} hold. 
    In addition, suppose $\phi$ has $r$ real roots and $c$ conjugate pairs ($r+2c = p$), and compute $G=2^rq^c$ if $q$ even and $G=q^c$ if $q$ odd.
    Then the identified set for $(\phi,v)$ contains at most $G$ points and can be computed as follows:
    \begin{enumerate}
        \item[1.] given $\phi$, find $\{\lambda_i\}_{i=1}^p$,\footnotemark
        \item[2.] for $g=1:G$:
        compute a candidate $\{\tilde\lambda^{(g)}_i\}_{i=1}^p$ using statements (iii) and (iv) of \cref{lemma_aliasingbound}, and check

        \begin{itemize}
            \item[(a)] (if stock) $D_i(\tilde\lambda^{(g)})/D_i(\lambda) = \kappa^{(g)}$ for $i=1:p$ and some $\kappa^{(g)} \in \reals_{++}$;
            \item[(b)] (if flow) $\sbr{D_i(\tilde\lambda^{(g)}) S_q(\lambda_i)} / \sbr{D_i(\lambda) S_q(\tilde\lambda_i^{(g)})} = \kappa^{(g)}$ for $i=1:p$ and some $\kappa^{(g)} \in \reals_{++}$;
            \item[(c)] (if flow) $V(\tilde\lambda^{(g)},\kappa^{(g)}v) = V(\lambda,v)$, 
        \end{itemize}
        where 
        $D_i(\lambda) = \prod_{j\neq i}(1-\lambda_j/\lambda_i) \prod_{j}(1-\lambda_i\lambda_j)$,
        $S_q(\lambda_i) = \sum_{|m|\leq q-1} (q-|m|)\lambda_i^m$,
        and $V(\lambda,v) = \sum_{i=1}^p K_i(\phi,v)\sbr{ q + 2\sum_{m=1}^{q-1}(q-m)\lambda_i^m }$.
        If $\{\tilde\lambda^{(g)}_i\}_{i=1}^p$ passes all checks, then $(\tilde\phi^{(g)},\kappa^{(g)}v)$ is in the identified set.
    \end{enumerate}
}

\footnotetext{
    The proof of \cref{lemma_rootsandpolynomials} shows how we can find $\{\lambda_i\}_{i=1}^p$ given $\phi$ and, in the end, $\tilde\phi^{(g)}$ given $\{\tilde\lambda^{(g)}_i\}_{i=1}^p$.
}

\cref{prop_ARpq_ID_computationofIDset} allows me to numerically verify that the previously discussed results generalize to the statements in \cref{conj_ARpq_stock_ID,conj_ARpq_flow_ID}.
For this purpose, a few implementation steps are required.
I summarize the resulting procedure as \cref{algo_verifyconj}.

\myalgo{Numerical Verification of \cref{conj_ARpq_stock_ID,conj_ARpq_flow_ID}}{algo_verifyconj}{
    Fix $P$, 
    $Q$,
    $v$,\footnote{
        This is without loss of generality; all checks in \cref{prop_ARpq_ID_computationofIDset} are invariant to $v$: the two ratios in checks (a) and (b) do not use $v$, and check (c) is homogeneous of degree one in $v$.
    }
    $\epsilon$,
    $\text{tol}$, and
    $M$.

    \noindent For $p=1:P$, draw $M$ valid $\phi^{(p,m)}$ via their root multisets $\{\lambda_{i}^{(p,m)}\}_{i=1}^p$ as follows:\footnote{
        Drawing roots rather than $\phi$ is more efficient, as the domains under which they satisfy assumptions are the same across $i$ and invariant to $p$.
    }
    \begin{enumerate}
        \item draw $c^{(p,m)}$ from $0:\text{floor}(p/2)$ with equal probabilities and set $r^{(m)}=p-2c^{(p,m)}$;
        \item draw $r^{(p,m)}$ real roots from $U(-1,1)$;
        \item draw $c^{(p,m)}$ complex roots via radius from $U(0,1)$ and angle from $U(0,\pi)$;
        \item verify \cref{ass_ARpq_ID_stationarity,ass_ARpq_ID_actualorderisp,ass_ARpq_ID_distincteigvalqthpowers} hold by a margin: 
        $$ \min_i |\lambda_{i}^{(p,m)}| > \epsilon \; , \quad \max_i |\lambda_{i}^{(p,m)}| < 1 - \epsilon \; , \quad \min_{i\neq j}|\br{\lambda_{i}^{(p,m)}}^q - \br{\lambda_{j}^{(p,m)}}^q| > \epsilon \; \text{for} \; q=1:Q \; . $$
    \end{enumerate}

    \noindent Let $\calH = \cbr{(p,q,m): \; p=1:P, \; q=1:Q, \; m=1:M}$.
    For each specification $h \in \calH$:
    \begin{enumerate}
        \item construct the $G(h)$ candidates $\{\tilde\lambda^{(g(h))}\}_{g(h)=1}^{G(h)}$ via \cref{lemma_aliasingbound}
        \item compute $D_i(\lambda^{(h)})$ and $S_q(\lambda_i^{(h)})$ for $i=1:p$ as well as $V^{(h)} = V(\tilde\lambda^{(h)},v)$
        \item for each candidate $g(h)$ compute\footnote{
            The implied ratios may be complex for some candidates.
            Since checks (a) and (b) in \cref{prop_ARpq_ID_computationofIDset} require $\kappa^{(g)} \in \reals_{++}$, the diagnostics $C^{(g(h))}_{s,1}$ and $C^{(g(h))}_{f,1}$ compare the ratios themselves -- not their moduli -- to the non-negative numbers $\kappa_s^{(g(h))}$ and $\kappa_f^{(g(h))}$, respectively.
            Comparing only moduli would not suffice: by statement (iii) in \cref{lemma_aliasingbound}, the candidate roots are conjugate-closed with the same index pairing as the true roots, hence $R^{(g(h))}_{s,i'} = \overline{R^{(g(h))}_{s,i}}$ and the moduli always coincide in pairs.
        } 
        \begin{equation*}
            \begin{alignedat}{2}
                R^{(g(h))}_{s,i}
                    &=
                    \frac{D_i(\tilde\lambda^{(g(h))})}{D_i(\lambda^{(h)})}
                    \; , \; i=1:p \; , \qquad
                &
                R^{(g(h))}_{f,i}
                    &=
                    \frac{D_i(\tilde\lambda^{(g(h))})S_q(\lambda_i^{(h)})}
                        {D_i(\lambda^{(h)})S_q(\tilde\lambda_i^{(g(h))})}
                    \; , \; i=1:p \; ,
                \\
                \kappa_s^{(g(h))}
                    &=
                    \max_{i=1:p}\bigl|R^{(g(h))}_{s,i}\bigr|
                    \; ,
                &
                \kappa_f^{(g(h))}
                    &=
                    \max_{i=1:p}\bigl|R^{(g(h))}_{f,i}\bigr|
                    \; , \qquad
                \\
                C^{(g(h))}_{s,1}
                    &=
                    \max_{i=1:p}
                    \frac{\bigl|R^{(g(h))}_{s,i}-\kappa_s^{(g(h))}\bigr|}
                        {\kappa_s^{(g(h))}}
                    \; , \qquad
                &
                C^{(g(h))}_{f,1}
                    &=
                    \max_{i=1:p}
                    \frac{\bigl|R^{(g(h))}_{f,i}-\kappa_f^{(g(h))}\bigr|}
                        {\kappa_f^{(g(h))}}
                    \; , 
                \\
                \tilde V^{(g(h))}
                    &=
                    V\bigl(\tilde\lambda^{(g(h))},\kappa_f^{(g(h))}v\bigr)
                    \; , \qquad
                &
                C^{(g(h))}_{f,2}
                    &=
                    \frac{\bigl|\tilde{V}^{(g(h))}-V^{(h)}\bigr|}
                        {\bigl|V^{(h)}\bigr|}
                    \; .
            \end{alignedat}
        \end{equation*}
        \item search for violations of \cref{conj_ARpq_stock_ID}: throw message if\footnote{
                The candidates $g(h)$ corresponding to $\phi^{(h)}$ and $\phi^{-(h)}$ can be identified by their roots: the roots of the former coincide with those of the true $\phi^{(h)}$, and the roots of the latter (if $q$ is even) are the sign-flipped roots of the true $\phi^{(h)}$.
            } 
        \begin{itemize}
            \item $C^{(g(h))}_{s,1} \geq \text{tol}$ for $g(h)$ corresponding to $\phi^{(h)}$; or 
            \item $C^{(g(h))}_{s,1} \geq \text{tol}$ for $g(h)$ corresponding to $\phi^{-(h)}$ and $q$ is even; or 
            \item $C^{(g(h))}_{s,1} < \text{tol}$ for some other $g(h)$. 
        \end{itemize}  
        \item search for violations of \cref{conj_ARpq_flow_ID}: throw message if
        \begin{itemize}
            \item $C^{(g(h))}_{f,1} \geq \text{tol}$ or $C^{(g(h))}_{f,2} \geq \text{tol}$ for $g(h)$ corresponding to $\phi^{(h)}$; or 
            \item $C^{(g(h))}_{f,1} < \text{tol}$ and $C^{(g(h))}_{f,2} < \text{tol}$ for some other $g(h)$. 
        \end{itemize}         
    \end{enumerate}
}

\begin{figure}[t]
	\begin{center}
		\vspace*{0pt}
		\begin{subfigure}[b]{\plotSizeDouble\textwidth}
			\subcaption*{\scriptsize{Check (a)-Diagnostic $C^{(g(h))}_{s,1}$, $q$ odd}}
				\centering
				\includegraphics[width=1\textwidth, clip]{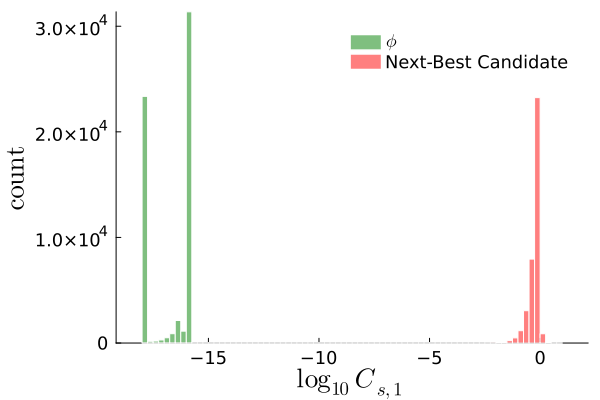}
		\end{subfigure}
		\begin{subfigure}[b]{\plotSizeDouble\textwidth}
			\subcaption*{\scriptsize{Check (a)-Diagnostic $C^{(g(h))}_{s,1}$, $q$ even}}
				\centering
				\includegraphics[width=1\textwidth, clip]{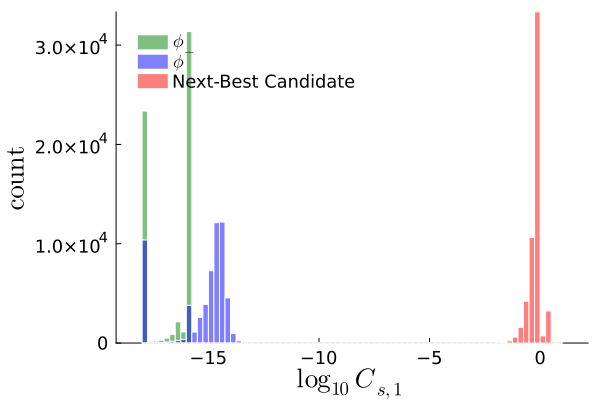}
		\end{subfigure}
		\vspace*{-10pt}
	\end{center}
    \caption{Numerical Verification of \cref{conj_ARpq_stock_ID} (Stock Case)\\[5pt]
    \scriptsize {\em Notes:} 
    The left plot shows $C^{(g(h))}_{s,1}$ across specifications $h$ with $q$ odd, the right plot shows this statistic across $h$ with $q$ even.
    The histogram in green shows the values for the respective candidate corresponding to the true $\phi^{(h)}$,
    the histogram in blue shows the values for the candidate corresponding to $\phi^{-(h)}$, and
    the histogram in red shows the minimal $C^{(g(h))}_{s,1}$ across all other candidates $g(h)$.
    The x-axis is in $\log_{10}$-scale and values lower than $10^{-18}$ are fixed at that value.}
    \label{plot_verifyconj_stock}

\end{figure}

\begin{figure}[t]
	\begin{center}
		\vspace*{0pt}
		\begin{subfigure}[b]{\plotSizeDouble\textwidth}
			\subcaption*{\scriptsize{Check (b)-Diagnostic $C^{(g(h))}_{f,1}$}}
				\centering
				\includegraphics[width=1\textwidth, clip]{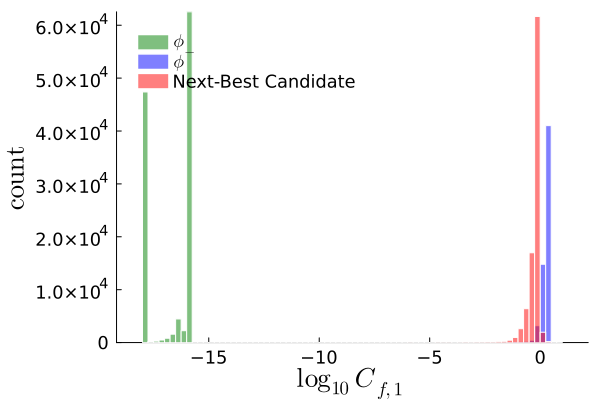}
		\end{subfigure}
		\begin{subfigure}[b]{\plotSizeDouble\textwidth}
			\subcaption*{\scriptsize{Check (c)-Diagnostic $C^{(g(h))}_{f,2}$}}
				\centering
				\includegraphics[width=1\textwidth, clip]{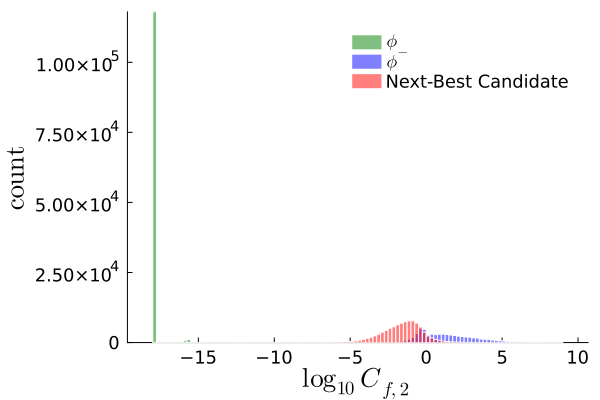}
		\end{subfigure}
		\vspace*{-10pt}
	\end{center}
    \caption{Numerical Verification of \cref{conj_ARpq_flow_ID} (Flow Case)\\[5pt]
    \scriptsize {\em Notes:} 
    The left plot shows $C^{(g(h))}_{f,1}$ and the right plot shows $C^{(g(h))}_{f,2}$ across specifications $h$.
    The histogram in green shows the values for candidates corresponding to the true $\phi^{(h)}$,
    the histogram in blue shows the values for candidates corresponding to $\phi^{-(h)}$, and
    the histogram in red shows the minimal value in the respective statistic across all other candidates $g(h)$.
    Note that the histogram in blue only uses specifications $h$ with $q$ even, as $\phi^{-(h)}$ is not a candidate for $q$ odd.
    The x-axis is in $\log_{10}$-scale and values lower than $10^{-18}$ are fixed at that value.}
    \label{plot_verifyconj_flow}

\end{figure}

I fix 
$P=12$, 
$Q=10$,
$v=1$,\footnote{
    This is without loss of generality; all checks in \cref{prop_ARpq_ID_computationofIDset} are invariant to $v$: the two ratios in checks (a) and (b) do not use $v$, and check (c) is homogeneous of degree one in $v$.
}
$\epsilon=0.01$,
$\text{tol}=10^{-8}$, and
$M=1000$.
I find no violations of \cref{conj_ARpq_stock_ID,conj_ARpq_flow_ID} in any of the $P \times Q \times M = 120,000$ specifications.
Furthermore, the candidates that are not in the identified sets stated in \cref{conj_ARpq_stock_ID,conj_ARpq_flow_ID} fail the required checks in \cref{algo_verifyconj} by a clear margin, as illustrated in \cref{plot_verifyconj_stock,plot_verifyconj_flow}.

\cref{plot_verifyconj_stock} illustrates the results of \cref{algo_verifyconj} for the stock case.
Across all specifications $h$ with $q$ odd, the left panel shows a histogram of the check(a)-diagnostic $C^{(g(h))}_{s,1}$ for the respective candidate corresponding to the true $\phi^{(h)}$ in green and the minimal $C^{(g(h))}_{s,1}$ across all other candidates $g(h)$ in red.\footnote{
    Note that the candidate corresponding to $\phi^-$ is not a candidate under $q$ odd, as then $\tilde\lambda_i^q \neq \lambda_i^q$ for $\tilde\lambda_i = - \lambda_i$.
}
The values on the x-axis are converted to $\log_{10}$-scale and censored at $10^{-18}$. 
In each specification, $C^{(g(h))}_{s,1} \approx 0$ for the candidate corresponding to $\phi^{(h)}$ -- indicating that this candidate is included in the identified set --, whereas this statistic clearly deviates from zero for the \myquote{next-best} candidate.
The right panel of \cref{plot_verifyconj_stock} shows the results for the stock case under $q$ even.
It shows $C^{(g(h))}_{s,1}$ for the candidate corresponding to the true $\phi^{(h)}$ in green,
$C^{(g(h))}_{s,1}$ for the candidate corresponding to $\phi^{-(h)}$ in blue, 
and the minimal $C^{(g(h))}_{s,1}$ across all other candidates $g(h)$ in red.
As required by \cref{conj_ARpq_stock_ID}, $C^{(g(h))}_{s,1} \approx 0$ for the candidates corresponding to $\phi^{(h)}$ and $\phi^{-(h)}$, with a noticeable gap to the value under the next-best candidate.

\cref{plot_verifyconj_flow} illustrates the results of \cref{algo_verifyconj} for the flow case with the same x-axis-scaling and -censoring as in \cref{plot_verifyconj_stock}.
The left panel depicts the check(b)-diagnostic $C^{(g(h))}_{f,1}$ for the candidate corresponding to the true $\phi^{(h)}$ in green and -- in case $q$ is even -- for the candidate corresponding to $\phi^{-(h)}$ in blue, and it shows the minimal $C^{(g(h))}_{f,1}$ across all other candidates $g(h)$ in red.
As required by \cref{conj_ARpq_flow_ID}, $C^{(g(h))}_{f,1} \approx 0$ for the candidate corresponding to $\phi^{(h)}$, with a noticeable gap to this statistic under other candidates.
The right panel shows analogous results for the check(c)-diagnostic $C^{(g(h))}_{f,2}$.





\section{Estimation}
\label{sec_est}




By \cref{prop_ARpq_stock_observeddynamics,prop_ARpq_flow_observeddynamics}, the observed process follows an ARMA whose autoregressive coefficients, moving-average coefficients and innovation variance -- collected in $\vartheta$ -- are smooth functions of $\theta=(\phi',v)'$ given $q$.
Under \cref{ass_ARpq_ID_stationarity,ass_ARpq_ID_actualorderisp,ass_ARpq_ID_distincteigvalqthpowers,ass_ARpq_ID_normality}, this ARMA is stationary, invertible and Gaussian (see \cref{sec_observeddynamics}), and the Jacobian $G(\theta) = \partial\vartheta/\partial\theta'$ has full column rank, since $\partial\rho_q/\partial\phi'$ is nonsingular when the $\lambda_i^q$ are distinct and non-zero.
Suppose in addition that $\theta$ lies in the interior of a compact subset of $\Theta_p$ on which it is point-identified -- which by \cref{conj_ARpq_stock_ID,conj_ARpq_flow_ID} requires excluding $\phi^-$ under stock sampling with even $q$ and holds otherwise under \cref{ass_ARpq_ID_stationarity,ass_ARpq_ID_actualorderisp,ass_ARpq_ID_distincteigvalqthpowers,ass_ARpq_ID_normality} -- and that the two polynomials of the observed ARMA share no root.
Then the Gaussian ML estimator $\hat\theta$ is consistent and
$$ \sqrt{T}\br{\hat\theta-\theta} \convdist N\br{0, \sbr{G(\theta)'\calI(\vartheta)G(\theta)}^{-1}} \; , $$
where $\calI(\vartheta)$ is the asymptotic information matrix of the unrestricted ARMA, block-diagonal in the ARMA-coefficients and the innovation variance \citep[Proposition 12.1 with Remarks 2 and 3, applied with $K=1$]{Luetkepohl2005}.\footnote{
    Proposition 12.1 is stated for linear restrictions on the ARMA-coefficients.
    Its Remark 2 extends it to any uniquely identified parametrization, which is what point-identification and the full rank of $G$ deliver here.
    The coefficient-block of $\calI(\vartheta)$ is given in closed form by \citet[Eq. (8.8.3)]{BrockwellDavis1991}, and the variance-block equals $1/(2\sigma^4)$, with $\sigma^2$ the innovation variance of the observed ARMA.
}
For $q=1$, $\vartheta=\theta$ and the result reduces to the standard asymptotics of the estimator $\hat\theta$ composed of $\hat\phi = \br{X'X}^{-1}X'Y$ and $\hat{v} = T^{-1}(Y-X\hat{\phi})'(Y-X\hat{\phi})$, where $Y$ is $T\times 1$ and stacks $y_t$ for $t=1:T$, and $X$ is $T \times p$ and stacks $x_t' = (y_{t-1}, ..., y_{t-p})$ for $t=1:T$.
For $q>1$, $G$ has to be computed numerically.

The restricted information matrix $G'\calI G$ is the one \citet[Eq. (A.3)]{PalmNijman1984} use for their efficiency comparisons, and \citet[Eq. (B.2)]{NijmanPalm1990b} state the same formula for a flow-sampled AR(1).\footnote{
    \citet[Appendix B]{JordaMarcellino2004} obtain consistency and asymptotic Normality of Kalman-filter ML for the parameters of the high-frequency process under a time-varying $q$, with global identifiability as the explicit condition.
}
These results presuppose point-identification, for which \cref{sec_AR_ID} provides the conditions.

\cref{ass_ARpq_ID_normality} matters only for the covariance matrix.
Without it, $\hat\theta$ is a quasi-ML estimator.
It remains consistent since the limit of the Gaussian likelihood depends on the autocovariances of $y_t$ alone.
It also remains asymptotically Normal if $e_\tau$ has finite fourth moments.
The $v$-block of $\sbr{G'\calI G}^{-1}$ then depends on the fourth moment of $e_\tau$, 
while its $\phi$-block remains valid if the innovations of the observed ARMA are i.i.d. \citep[Remark 4 to Proposition 12.1]{Luetkepohl2005}.
This holds under $q=1$ or for a stock-sampled AR(1), whose observed innovations are sums over disjoint blocks of $e_\tau$'s.
For $p\geq2$ or flow sampling, however, consecutive observed innovations are built from overlapping blocks of $e_\tau$ and therefore not independent.


\paragraph*{Precision}

How precisely $\phi$ can be estimated depends on $q$, and it does so non-monotonically.
\cref{appsec_est} shows this analytically for the simplest case, an AR($1,q$) observed as a stock variable.
The observed process is then an AR($1$) with autocorrelation coefficient $\psi = \phi^q$ and the per-observation information about $\phi$ equals $\calI_{\phi\phi}(q) = q^2\phi^{2q-2}/(1-\phi^{2q})$.
For a persistent process, information about $\phi$ first rises with $q$ up to $q^{*} \approx -0.797/\log(|\phi|)$.
Considering that $q \in \naturals$, information is decreasing throughout only if $|\phi| < 0.58$.
Treating $q$ as continuous, the optimum is $q^*=0.66,1.15,2.23,7.56,15.53$ for $|\phi|=0.3,0.5,0.7,0.9,0.95$, respectively, so that the optimal integer $q$ is $1,1,2,8,16$.

\citet[Eq. (22)]{NijmanPalm1990b} derive the asymptotic variance of $\hat\phi$ for this model by the same delta-method step, and at $q=2$ it yields the relative efficiency $(1+\phi^2)/(2\phi^2)$ of the complete-data estimator reported by \citet[p.~1427]{PalmNijman1984} and \citet[p.~339]{PierseSnell1995}.
They find that imperfect observation always loses information because they hold the number of high-frequency periods fixed, in which case the number of actual observations falls in proportion to $q$, and information about $\phi$ per high-frequency period is then $\calI_{\phi\phi}(q)/q$, which decreases in $q$ for any $\phi$.
Instead, I hold the number of observations $T$ fixed.
This normalization is the relevant one when a given sample is modeled at alternative underlying frequencies: the number of observations is fixed and $q$ indexes the model. %
In that case we obtain the non-monotonicity and the possible optimum $q^*>1$ discussed above, which, to my knowledge, are new to the literature.


\section{Conclusion}
\label{sec_conclusion}

In this paper, I derive the observed dynamics, characterize identification and discuss inference for regularly but infrequently observed or temporally aggregated autoregressions.
The key to my generalization of earlier results in the literature is my explicit representation of the observed process in terms of the autoregressive coefficients in the underlying high-frequency model.

It would be interesting to apply this representation approach to more general underlying ARMA processes, as studied in \citetwo{Brewer1973}{PalmNijman1984}.
The resulting identification analysis would be more complex than mine, as under MA-errors the finiteness of the identified set is no longer guaranteed.
For example, an MA(1) process $x_\tau = e_\tau + \theta e_{\tau-1}$, $e_\tau \sim WN(0,v)$ observed every second period (stock case) yields a White Noise with variance $v(1+\theta^2)$ as the observed process and hence a whole continuum of parameter values observationally equivalent to the true $(\theta,v)$ \citep[pp. 1418--1419]{PalmNijman1984}.


\begin{spacing}{1.0}
	\bibliography{../../P017}
\end{spacing}


\newpage

\renewcommand{\thepage}{A.\arabic{page}}
\setcounter{page}{1}

\begin{appendix}
	\markright{This Version: \today }
	\renewcommand{\theequation}{A.\arabic{equation}}
	\setcounter{equation}{0}

	\renewcommand*\thetable{A-\arabic{table}}
	\setcounter{table}{0}
	\renewcommand*\thefigure{A-\arabic{figure}}
	\setcounter{figure}{0}

	\addtocontents{toc}{\protect\setcounter{tocdepth}{1}}


\section{Representation}
\label{appsec_observeddynamics}

\myproofof{lemma_PsiqRewriting}{
    To see that such a $\rho_q$ exists, note first that we have $\Psi_q(\wq z) = \Psi_q(z)$ because $\wq^q = 1$. 
    If we write $\Psi_q(z) = \sum_{l=0}^{pq} b_l z^l$, then $\Psi_q(\wq z) = \sum_{l=0}^{pq} b_l \wq^l z^l$ and so we must have $b_l (\wq^l -1) = 0$, which by $\wq^q=1$ implies that $b_l = 0$ for each $l$ not divisible by $q$.

    For $q=1$, we trivially have $\rho_1 = \phi$ with real coefficients.
    For $q\geq 2$, it suffices to show $\prod_{j=0}^{q-1}(1-\lambda \wq^j z) = 1 -
    \lambda^q z^q$ for any $\lambda \in \complexnumbers$.
    The statement then follows by evaluating at $\lambda = \lambda_i$ and taking the product over $i$. 
    For $\lambda = 0$, the claims trivially hold. 
    For $\lambda \neq 0$,
    both sides are polynomials in $z$ of degree $q$ with the same $q$ distinct roots $\{\wq^{-j}\lambda^{-1}\}_{j=0}^{q-1}$, and are therefore equal up to a multiplicative constant.
    Since both have constant term $1$, that constant is one: the two polynomials are equal. 

    To see that $\rho_q$ has real coefficients, note that because $\phi$ is real, non-real roots come in conjugate pairs, and so $\{\lambda_i\}$ is closed under
    complex conjugation: $\{\bar \lambda_1, ..., \bar \lambda_p\} = \{\lambda_1, ..., \lambda_p\}$.
    Because $\bar{\lambda^q} = (\bar{\lambda})^q$, the same holds for $\{\mu_i\}$, $\mu_i = \lambda_i^q$.
    We can write 
    $$ \rho_q(z) = \sum_{k=0}^p (-1)^k e_k(\mu_1,...,\mu_p)z^k \; , \quad e_k(\mu) = \sum_{i_1<...<i_k}\mu_{i_1} \cdot ... \cdot \mu_{i_k} \; . $$
    Each $e_k(\mu)$ is real because $e_k(\mu)=e_k(\bar\mu)$: 
    since conjugation distributes over sums and products, $\bar{e_k(\mu)}=e_k(\bar\mu)$,
    and $\{\mu_i\}$ is closed under complex conjugation and $e_k(\mu)$ is invariant under permutation of its arguments, $e_k(\mu) = e_k(\bar\mu)$.
    $\blacksquare$
}


\section{Identification}
\label{appsec_ID}




\subsection{Supporting Lemmas}
\label{appsubsec_ID_lemmas}

\mylemma{}{lemma_reflectionisdistinct}{
    Let $q\in \naturals$ be even and suppose \cref{ass_ARpq_ID_actualorderisp,ass_ARpq_ID_distincteigvalqthpowers} hold.
    Then $\phi^- \neq \phi$.
}

\myproof{
    $\phi^- = \phi$ would require $\phi_j = 0$ for all odd $j$.
    If $p$ is odd, this contradicts \cref{ass_ARpq_ID_actualorderisp}.
    If $p$ is even, then $\phi(z)=\phi^-(z)=\phi(-z)$ implies that the roots must come in $\pm$ pairs, i.e. $\lambda_i = -\lambda_j$ for some $i\neq j$.
    For $q$ even, this yields $\lambda_i^q = \lambda_j^q$, contradicting \cref{ass_ARpq_ID_distincteigvalqthpowers}.
    $\blacksquare$
}

\mylemma{}{lemma_phiCoincidesIfPhiCoincides}{
    Let $\tilde\phi$ and $\hat\phi$ be real polynomials with degrees $\le p$ and constant term $1$, and suppose they satisfy \cref{ass_ARpq_ID_stationarity}.
    Let $\tilde\Phi(z) =\tilde\phi(z)\tilde\phi(z^{-1})$, and likewise for $\hat\Phi$.
    If $\tilde\Phi = \hat\Phi$, then $\tilde\phi = \hat\phi$.
}

\myproof{
    Write $\tilde\phi(z) = \prod_{i=1}^{\tilde d}(1-\tilde\lambda_iz)$ with $\tilde d = \deg\tilde\phi$ and all $\tilde\lambda_i \neq 0$, $|\tilde\lambda_i|<1$.
    The zeros of $\tilde\Phi(z) = \tilde\phi(z)\tilde\phi(z^{-1})$ in $\complexnumbers\setminus\{0\}$ are the list $(1/\tilde\lambda_1,\dots,1/\tilde\lambda_{\tilde d},\tilde\lambda_1,\dots,\tilde\lambda_{\tilde d})$ (counted with multiplicity).
    Note $1/\tilde\lambda_i \neq \tilde\lambda_j$ for all $i,j$, since $|\tilde\lambda_i\tilde\lambda_j|<1$. 
    Also, since all $|\tilde\lambda_i|<1$, the entries with modulus larger than one are exactly the reciprocals $1/\tilde\lambda_i$.
    Analogous statements apply for $\hat\phi$, $\hat\Phi$ and the list $(1/\hat\lambda_1,\dots,1/\hat\lambda_{\hat d},\hat\lambda_1,\dots,\hat\lambda_{\hat d})$.
    The equality $\tilde\Phi = \hat\Phi$ implies that the two lists coincide (up to ordering; including multiplicities).
    In particular, their sublists of entries with modulus larger than one coincide, i.e.$(1/\tilde\lambda_1,\dots,1/\tilde\lambda_{\tilde d})$ and $(1/\hat\lambda_1,\dots,1/\hat\lambda_{\hat d})$ coincide up to ordering.
    Hence, $\tilde d = \hat d$, the eigenvalue lists coincide (up to ordering), and $\tilde\phi=\hat\phi$.
    $\blacksquare$
}

\mylemma{}{lemma_ACF_ARp_candidate}{
    Consider $(\tilde\phi,\tilde v) \in\Theta_p$ with effective order $\tilde p = \max\{j:\tilde\phi_j\neq0\}$ and distinct eigenvalues $\tilde\lambda_1,\dots,\tilde\lambda_{\tilde d}$ with multiplicities $\tilde m_1,\dots,\tilde m_{\tilde d}$.\footnote{
        The multiplicity $\tilde{m}_i$ is the number of times the eigenvalue $\tilde \lambda_i$ is repeated as a root, i.e. we have $\tilde \phi(z) = \prod_{i=1}^{\tilde d} (1-\tilde \lambda_i z)^{\tilde m_i}$ and $\sum_{i=1}^{\tilde d} \tilde m_i = \tilde p$.
        Also, note that by construction $\tilde\lambda_i\neq0 \; \forall \; i$ because $\tilde\phi_{\tilde p}\neq 0$ and $\tilde\phi_p = (-1)^{p+1}\prod_i \tilde\lambda_i$.
    } 
    If $\tilde p = 0$ (White Noise), then $\gamma_{\tilde x}(h) = \tilde v \one{h=0}$.
    If $\tilde p \geq 1$,     
    then there exist polynomials $\tilde P_i$ with $\deg \tilde P_i \le \tilde m_i - 1$ s.t.
    \[
        \gamma_{\tilde x}(h) = \sum_{i=1}^{\tilde d} \tilde P_i(h) \tilde\lambda_i^{h} \quad \text{for} \quad h \ge 0 \; .
    \]
}

\myproof{
    The statement for $\tilde p= 0$ can be verified trivially. For the statement for $\tilde p \geq 1$, see \citet[Section 3.3]{BrockwellDavis1991}. $\blacksquare$ 
}

\mylemma{}{lemma_ACF_ARp_truevalue}{
    Let $(\phi,v) \in \Theta_p$ satisfy \cref{ass_ARpq_ID_actualorderisp} and have $p$ distinct roots $\lambda_1,\dots,\lambda_p$. 
    Then
    $$ 
    \gamma_x(h) = \sum_{i=1}^{p} K_i(\phi,v)\lambda_i^h \; \; \text{for} \; h \geq 0 \; , \quad 
    K_i(\phi,v) =  \frac{v}{ \prod_{j\neq i}\br{ 1 - \lambda_j/\lambda_i }\;
    \prod_{j=1}^{p}\br{ 1-\lambda_i\lambda_j } } \neq 0 \; . $$
}

\myproof{
    Note that the stated constants $K_i$ are well-defined and non-zero: $1-\lambda_j/\lambda_i \neq 0$ by distinctness and
    $1-\lambda_i\lambda_j \neq 0$ by $|\lambda_i\lambda_j| < 1$ (\cref{ass_ARpq_ID_stationarity}).
    Define
    $$ H(z) = \sum_{i=1}^p K_i\left[\frac{1}{1-\lambda_i z} +
    \frac{\lambda_i z^{-1}}{1-\lambda_i z^{-1}}\right] \; .$$
    I first show that $\Gamma_x(z) = H(z)$ for the two rational functions $\Gamma_x(z)$ and $H(z)$.
    For this purpose, consider $R(z) = \Gamma_x(z) - H(z)$, and recall that $\Gamma_x(z) = v/\Phi(z)$ with $\Phi(z)=\phi(z)\phi(z^{-1})$.
    First, note that $R$ has no poles in $\complexnumbers\setminus\{0\}$.
    Its only candidate poles in $\complexnumbers\setminus\{0\}$ are the $2p$ distinct points $z = 1/\lambda_i$ and $z = \lambda_i$, $i = 1:p$.\footnote{
        They are distinct because $1/\lambda_i \neq 1/\lambda_j$ and $\lambda_i \neq \lambda_j$ by distinctness of $\lambda_i$, $i=1:p$, 
        and because $1/\lambda_i \neq \lambda_j$ as $|\lambda_i\lambda_j| < 1$ by \cref{ass_ARpq_ID_stationarity}.
    }
    We have $\lim_{z \to 1/\lambda_i} \Gamma_x(z) (1-\lambda_i z) = K_i$
    because 
    $\lim_{z \to 1/\lambda_i} \phi(z)/(1-\lambda_i z) = \prod_{j\neq i}(1-\lambda_j/\lambda_i)$ 
    and $\lim_{z \to 1/\lambda_i} \phi(z^{-1}) = \phi(\lambda_i) = \prod_j(1-\lambda_j\lambda_i)$.
    Also, $\lim_{z \to 1/\lambda_i}H(z)(1-\lambda_i z) = K_i$ (from the first term of the $i$th summand, while all other summands are annihilated by the factor $(1-\lambda_i z) \to 0$). 
    Analogously, $\lim_{z\to\lambda_i}\Gamma_x(z)(1-\lambda_i z^{-1}) = K_i$ because 
    $\lim_{z \to \lambda_i} \phi(z) = \phi(\lambda_i)$ and $\lim_{z \to \lambda_i} \phi(z^{-1})/(1-\lambda_i z^{-1}) = \prod_{j\neq i}(1-\lambda_j/\lambda_i)$, 
    and $\lim_{z \to \lambda_i}H(z)(1-\lambda_i z^{-1})=K_i$ (from the second term of the $i$th summand).
    Hence, $R$ has no poles at $1/\lambda_i$ nor at $\lambda_i$, $i=1:p$.
    Second, as $z \to \infty$, $v/\Phi(z) = v z^p / [\prod_j(1-\lambda_j z)\prod_j(z-\lambda_j)] \to
    0$ and $H(z) \to 0$.
    Third, as $z \to 0$, both terms tend to $0$ as well. 
    A rational function with no poles in $\complexnumbers\setminus\{0\}$, no pole at $0$, and limit $0$ at $\infty$ is identically $0$. 
    Hence, $R=0$, and so $\Gamma_x = H$.

    Recall that $\Gamma_x(z) = \sum_{h=-\infty}^{\infty} \gamma_x(h) z^h$.
    By \cref{ass_ARpq_ID_stationarity}, 
    $1/(1-\lambda_i z) = \sum_{h\geq 0}\lambda_i^h z^h$ 
    and $\lambda_i z^{-1}/(1-\lambda_i z^{-1}) = \sum_{h \geq 1}\lambda_i^h z^{-h}$.
    Using these expressions and reading off the coefficients of $z^h$, $h\geq 0$ yields the expression for $\gamma_x(h)$ in the statement.
    $\blacksquare$
}

\mylemma{}{lemma_unique_expopolysums}{
    Let $\nu_1,\dots,\nu_K$ be distinct and non-zero scalars,
    let $Q_1,\dots,Q_K$ be polynomials, 
    and suppose $\exists \; h_0 \in \integers_+$ s.t. $\sum_{k=1}^K Q_k(h)\nu_k^h = 0$ for $h\ge h_0$. 
    Then $Q_k = 0$ for each $k$.
}

\myproof{
    For $h_0 =0$, see \citet[Section 3.6]{BrockwellDavis1991}. 
    For any $h_0 > 0$, define $Q_k'(h) = \nu_k^{h_0}Q_k(h+h_0)$, write $\sum_kQ_k'(h)\nu_k^h = 0$ for $h \ge 0$ and conclude that $Q_k' = 0$.
    Since $\nu_k \neq 0$, we must have $Q_k = 0$.
    $\blacksquare$


}

\mylemma{}{lemma_rootsandpolynomials}{
    Let $\calL_p$ be the collection of multisets $\lambda = \{\lambda_1, ..., \lambda_p\}$ with $\lambda_i \in \complexnumbers$, $\lambda_i \neq 0$, $i=1:p$ that are closed under complex conjugation.
    Let $\calP_p$ be the set of real polynomials $\phi(z) = -\sum_{j=0}^p \phi_j z^j$ with degree $p$ and constant term $\phi(0)=-\phi_0=1$.
    Then $T: \calL_p \rightarrow \calP_p$, $T(\lambda)(z) = \prod_{i=1}^p (1-\lambda_i z)$ is a bijection.
}

\myproof{
    Note that $T$ maps $\calL_p$ into $\calP_p$:
    given $\lambda \in \calL_p$, expanding the product yields
    $$ T(\lambda)(z) = \sum_{k=0}^{p}(-1)^k e_k(\lambda_1,\dots,\lambda_p) z^k \quad \text{with} \quad e_0(\lambda) = 1 \; , \quad e_k(\lambda) = \sum_{i_1<\dots<i_k} \lambda_{i_1}\cdots\lambda_{i_k} \; , \ k \geq 1 \; , $$
    i.e.\ $\phi_k = (-1)^{k+1}e_k(\lambda)$.
    Each $e_k(\lambda)$ is real by the same argument as in
    the proof of \cref{lemma_PsiqRewriting}, since $\lambda$ is closed
    under complex conjugation. 
    The constant term is $1$, and the degree is exactly $p$, since the coefficient of $z^p$ is $(-1)^p e_p(\lambda) = (-1)^p \prod_{i=1}^p \lambda_i \neq 0$.
    Hence $T(\lambda) \in \calP_p$.

    Now, let $\phi \in \calP_p$. 
    By the fundamental theorem of algebra, $\phi$ has exactly $p$ zeros $z_1, \dots, z_p$ in $\complexnumbers$, counted with multiplicity. 
    They are non-zero because $\phi(0) = 1 \neq 0$. 
    The multiset $\{z_i\}$ is closed under complex conjugation: since $\phi$ has real coefficients,
    $\overline{\phi(\bar z)} = \phi(z)$ for all $z$, so $z$ and $\bar
    z$ are zeros of $\phi$ of the same multiplicity. 
    Hence $\lambda = \{1/z_1,\dots,1/z_p\} \in \calL_p$: its elements are well-defined
    and non-zero, and it is closed under conjugation because $\overline{1/z} = 1/\overline{z}$. 
    The polynomials $T(\lambda)$ and $\phi$ both have degree $p$ and the same zero multiset $\{z_i\}$, so they are equal up to a multiplicative constant, and
    since both have constant term $1$, $T(\lambda) = \phi$.
    This proves surjectivity.
    $T$ is also injective: the factorized form shows that the zero multiset of $T(\lambda)$ is exactly
    $\{1/\lambda_1,\dots,1/\lambda_p\}$, so $\lambda$ is recovered from $T(\lambda)$ as the multiset of reciprocals of its zeros; two multisets with the same image under $T$ therefore coincide.

    The above shows how to compute the polynomial $T(\lambda)$ given $\lambda$.
    Conversely, given $\phi$, we can compute $T^{-1}(\phi)$ as reciprocals of the zeros of $\phi$ or, equivalently, as zeros of the reciprocal polynomial $z^p\phi(1/z) = z^p - \phi_1 z^{p-1} - \dots - \phi_p$, i.e. eigenvalues of its companion form-matrix.
    $\blacksquare$
}

\lemmamarker{lemma_aliasingbound}

\myproofof{lemma_aliasingbound}{
    Write $\mu_i = \lambda_i^q$ for $i = 1:p$.
    By \cref{ass_ARpq_ID_actualorderisp,ass_ARpq_ID_distincteigvalqthpowers}, the $\mu_i$ are distinct and non-zero.
    Since $\tilde\lambda_i^q = \mu_i$ (re-indexed), $\tilde\phi$ also satisfies \cref{ass_ARpq_ID_distincteigvalqthpowers}, and we can write $\tilde\lambda_i = \wq^{k_i}\lambda_i$ for some $k_i \in \{0,1,...,q-1\}$.
    This $k_i$ is unique because $\wq^0,\dots,\wq^{q-1}$ are distinct.
    Thus, $\tilde\phi$ also satisfies \cref{ass_ARpq_ID_stationarity,ass_ARpq_ID_actualorderisp}: 
    its roots are $|\tilde\lambda_i| = |\wq^{k_i}\lambda_i| = |\wq^{k_i}||\lambda_i| = |\lambda_i| < 1$ by \cref{ass_ARpq_ID_stationarity},
    and they are non-zero because $\wq^{k_i} \neq 0$ for any $k_i$ and because $\lambda_i \neq 0$ by \cref{ass_ARpq_ID_actualorderisp}.
    Since all $\mu_i$ and all $\tilde\lambda_i$ are distinct, respectively, we can assign a unique $q$th power from $\{\mu_1,\dots,\mu_p\}$ to each root of $\tilde\phi$, and vice versa, and so we can uniquely index each root of $\tilde\phi$ by the index of its image. 
    This proves statements (i) and (ii).

    To show (iii) and (iv), I first show that conjugation permutes the candidate roots through the same index pairing $i \leftrightarrow i'$ as the true roots.
    Recall that since $\phi$ and $\tilde\phi$ have real coefficients, both $\{\lambda_i\}$ and $\{\tilde\lambda_i\}$ are closed under complex conjugation, i.e. for each root $\lambda_i$ of $\phi$, $\overline{\lambda_i}$ is also a root of $\phi$, and for each root $\tilde\lambda_i$ of $\tilde\phi$, $\overline{\tilde\lambda_i}$ is also a root of $\tilde\phi$.
    Also, recall that since all $\mu_i$ and all $\tilde\lambda_i$ are distinct, respectively, we can assign a unique $q$th power from $\{\mu_1,\dots,\mu_p\}$ to each root of $\tilde\phi$, and vice versa, and note that the same holds for roots of $\phi$.
    For fixed $i$, let $i'$ be the index with $\mu_{i'} = \overline{\mu_i}$.
    This $i'$ exists because $\overline{\mu_i} = \overline{\lambda_i^q} = (\overline{\lambda_i})^q$ is the $q$th power of the root $\overline{\lambda_i}$, 
    and it is unique because the $\mu_i$ are distinct.
    By the bijective mapping from roots to $q$th powers, we also have $\lambda_{i'} = \overline{\lambda_i}$ at that $i'$.
    Now consider the candidate root $\overline{\tilde\lambda_i}$.
    Its $q$th power is $\br{\overline{\tilde\lambda_i}}^q = \overline{\tilde\lambda_i^{ q}} = \overline{\mu_i} = \mu_{i'}$.
    By the bijective mapping from roots to $q$th powers, we have $\overline{\tilde\lambda_i} = \tilde\lambda_{i'}$, and thus also $\tilde\lambda_{i'} = \overline{\tilde\lambda_i}$.
    %

    Applying the above at some $i' \neq i$ (i.e. at some $i$ for which $\lambda_i \notin \reals$) directly yields statement (iii).
    If instead $\lambda_i \in \reals$, then $\mu_i \in \reals$, and so $i' = i$.
    Then, $\overline{\tilde\lambda_i} = \tilde\lambda_i$, i.e. $\tilde\lambda_i \in \reals$.
    This means that $\wq^{k_i} = \tilde\lambda_i/\lambda_i$ is a real $q$th root of unity. Hence, $\wq^{k_i} = 1$ if $q$ is odd and $\wq^{k_i} \in \{1,-1\}$ if $q$ is even.
    This proves statement (iv).

    Statement (v) on the number of root multisets $\{\tilde\lambda_i\}_{i=1}^p$ is obtained as follows.
    By (iv), every real root of $\phi$ leads to a single $\tilde\lambda_i=\lambda_i$ if $q$ is odd and two $\tilde\lambda_i=\pm \lambda_i$ if $q$ is even.
    By (ii) and (iii), every conjugate pair $\{\lambda_i, \lambda_{i'}\}$ admits $q$ choices, indexed by $k_i \in \{0,\dots,q-1\}$: $\tilde\lambda_i = \omega_q^{k_i}\lambda_i$, which pins down the second root as $\tilde\lambda_{i'} = \overline{\tilde\lambda_i}$.
    Hence there are $2^r q^c$ construction choices for $q$ even and $q^c$ for $q$ odd. 
    Distinct choices lead to distinct multisets: by construction, every element of such a multiset satisfies $\tilde\lambda_i^q = \mu_i$ for its index $i$, and since the $\mu_i$ are distinct, each element determines its index through its $q$th power.
    Thus, two such multisets can coincide only if they coincide index by index, and at any given index, distinct choices produce distinct roots, because $\omega_q^0,\dots,\omega_q^{q-1}$ are distinct and $\lambda_i \neq 0$.
    Finally, every multiset $\{\tilde\lambda_i\}_{i=1}^p$ consists of $p$ non-zero elements and is closed under complex conjugation.
    By \cref{lemma_rootsandpolynomials}, each such root multiset translates into a unique real polynomial with degree $p$ and constant term 1.
    $\blacksquare$
}




\subsection{Proofs of Propositions}
\label{appsubsec_ID_proofs}

\propositionmarker{prop_ARpq_ID_q1}

\myproofof{prop_ARpq_ID_q1}{
    If $(\tilde\phi,\tilde v) \in \Theta_p$ is observationally equivalent to $(\phi,v)$, then $\Gamma_{\tilde y} = \Gamma_y$, i.e. $\tilde v/\tilde\Phi(z) = v/\Phi(z)$ and therefore $\tilde\Phi(z) = c\Phi(z)$ with $c = \tilde v/v > 0$. 
    This means that the zero multisets of $\tilde\Phi$ and $\Phi$ in $\complexnumbers\setminus\{0\}$ coincide. 
    Together with \cref{ass_ARpq_ID_stationarity} this implies $\tilde\phi = \phi$, just as in the proof of \cref{lemma_phiCoincidesIfPhiCoincides}.
    Thus, $\tilde\Phi = \Phi$, and so $\tilde v = v$.
    $\blacksquare$
}

\propositionmarker{prop_ARpq_stock_observeddynamics_ID}

\myproofof{prop_ARpq_stock_observeddynamics_ID}{
    Write $\mu_i = \lambda_i^q$. 
    By \cref{ass_ARpq_ID_actualorderisp,ass_ARpq_ID_distincteigvalqthpowers}, the $\mu_i$ are distinct and non-zero. 
    Since $\gamma_y(h) = \gamma_x(qh)$, \cref{lemma_ACF_ARp_truevalue} implies 
    $\gamma_y(h) = \sum_{i=1}^{p} K_i \mu_i^h$ for $h \geq 0$ with $K_i = K_i(\phi,v) \neq 0$.
    Observational equivalence implies $\gamma_{\tilde y}(h) = \gamma_y(h)$
    for all $h \geq 0$.

    The candidate must have effective order $\tilde p \geq 1$.\footnote{
        If $\tilde p = 0$, \cref{lemma_ACF_ARp_candidate} gives $\gamma_{\tilde y}(h) = \gamma_{\tilde x}(qh) = \tilde v \mathbf{1}\{h=0\}$, so $\sum_{i=1}^p K_i\mu_i^h = 0$ for $h \geq 1$, and \cref{lemma_unique_expopolysums} yields $K_i = 0$ for all $i$, a contradiction.
    } 
    So, by \cref{lemma_ACF_ARp_candidate}, $\gamma_{\tilde y}(h) = \gamma_{\tilde x}(qh) = \sum_{i=1}^{\tilde d} \tilde P_i(qh)\br{ \tilde\lambda_i^q }^h$ for $h \geq 0$. 
    Let $\nu_1,\dots,\nu_K$ be the distinct values among $\tilde\lambda_i^q$, $i = 1:\tilde d$, and collect terms:
    $\gamma_{\tilde y}(h) = \sum_{k=1}^{K}\tilde Q_k(h)\nu_k^h$ with polynomials $\tilde Q_k$ and $K \leq \tilde d \leq \tilde p \leq p$.
    Note that all $\nu_k$ are non-zero, as $\tilde\lambda_i \neq 0$ for $i=1:\tilde{d}$ (see footnote in \cref{lemma_ACF_ARp_candidate}).

    Let $w_1,\dots,w_M$ be the distinct elements of
    $\{\mu_1,\dots,\mu_p\} \cup \{\nu_1,\dots,\nu_K\}$, 
    and define $R_m(h) = K_i\mathbf{1}\{w_m = \mu_i\} - \tilde Q_k(h)\mathbf{1}\{w_m = \nu_k\}$ s.t. $$ \sum_{m=1}^{M} R_m(h)w_m^h = \sum_{i=1}^p K_i\mu_i^h - \sum_{k=1}^K \tilde Q_k(h)\nu_k^h = 0 \; \; \text{for} \; h \geq 0\;. $$
    Since the $w_m$ are distinct and non-zero, \cref{lemma_unique_expopolysums} gives $R_m = 0$ for each $m$. 
    This shows, that every $\mu_i$ must appear in $\{\nu_k\}$; if $w_m = \mu_i$ for some $m$ and
    $i$ but $\mu_i \notin \{\nu_k\}$, then $R_m = K_i$, but $K_i = 0$ is a
    contradiction. 
    Hence $K \geq p$, and so $K = \tilde d = \tilde p = p$. 
    Consequently, $\{\tilde\lambda_i^q\} = \{\nu_k\} = \{\mu_i\} = \{\lambda_i^q\}$.
    By \cref{lemma_aliasingbound}, we can uniquely index the candidate roots s.t. $\tilde\lambda_i^q = \lambda_i^q$ for $i=1:p$.

    Then, applying \cref{lemma_ACF_ARp_truevalue} to $(\tilde\phi,\tilde v)$, we can see that $\tilde Q_k$ must be constants.
    With roots re-indexed s.t. $\tilde\lambda_i^q = \lambda_i^q$, the constants must match: $R_m \equiv 0$ gives $K_i(\tilde\phi,\tilde v) = K_i(\phi,v)$.

    Finally, by \cref{lemma_PsiqRewriting} and with re-indexed roots, $\tilde\Psi_q(z) = \prod_{i=1}^p\br{ 1 - \tilde\lambda_i^q z^q } = \prod_{i=1}^p (1-\mu_i z^q) = \Psi_q(z)$, and so $\tilde\rho_q = \rho_q$. 
    Also, because $\rho_q(L)y_t = u_t$ by \cref{prop_ARpq_stock_observeddynamics}, we have $\Gamma_u(w) = \rho_q(w)\rho_q(w^{-1})\Gamma_y(w)$, and the same identity holds with tildes.
    Since $\tilde \rho_q = \rho_q$ and $\Gamma_{\tilde y} = \Gamma_y$, we have $\Gamma_{\tilde u} = \Gamma_u$.
    $\blacksquare$
}

\propositionmarker{prop_ARpq_stock_ID}

\propositionpartmarker{prop_ARpq_stock_ID_generalcase}

\propositionpartmarker{prop_ARpq_stock_ID_qEven_reflectionequiv}

\propositionpartmarker{prop_ARp2_stock_ID}

\propositionpartmarker{prop_ARpq_stock_ID_realroots}

\myproofof{prop_ARpq_stock_ID_generalcase}{
    \cref{prop_ARpq_stock_observeddynamics_ID} shows that any observationally equivalent $(\tilde\phi,\tilde v)$ is in this set.
    Conversely, let $(\tilde\phi,\tilde v) \in \Theta_p$ be in the stated set.
    Then, by \cref{lemma_aliasingbound}, it satisfies \cref{ass_ARpq_ID_actualorderisp,ass_ARpq_ID_distincteigvalqthpowers}, and therefore has non-zero and distinct roots.
    In turn, applying \cref{lemma_ACF_ARp_truevalue} to both parameter points shows that $(\tilde\phi,\tilde v)$ implies the same observed autocovariances as $(\phi,v)$:
    $$ \gamma_{\tilde y}(h) = \sum_i K_i(\tilde\phi,\tilde v)\mu_i^h = \sum_i K_i(\phi,v)\mu_i^h = \gamma_y(h) \quad \forall \; h \geq 0\;. $$
    Because $y$ is Normal --- implied by \cref{ass_ARpq_ID_normality} --- and has mean zero,
    this corresponds to observational equivalence.
    $\blacksquare$
}


\myproofof{prop_ARpq_stock_ID_qEven_reflectionequiv}{
    Note that $\phi^{-}$ satisfies \cref{ass_ARpq_ID_stationarity}, and so $(\phi^-,v) \in \Theta_p$.
    Under $(\phi^-,v)$, the latent ACGF satisfies $\Gamma_{x^-}(z) = v/\br{ \phi(-z)\phi(-z^{-1}) } = \Gamma_x(-z)$, so $\gamma_{x^-}(h) = (-1)^h\gamma_x(h)$ and
    $\gamma_{y^-}(h) = (-1)^{qh}\gamma_x(qh) = \gamma_y(h)$ for $q$ even.
    By \cref{ass_ARpq_ID_normality}, this implies observational equivalence.
    $\blacksquare$
}


\myproofof{prop_ARp2_stock_ID}{
    Note that \cref{lemma_reflectionisdistinct} shows that the two points are distinct under $q$ even
    and \cref{prop_ARpq_stock_ID_qEven_reflectionequiv} shows that $(\phi^-,v)$ is observationally equivalent to $(\phi,v)$ under $q$ even.
    It remains to show that any $(\tilde\phi,\tilde v) \in \Theta_p$ that is observationally equivalent to $(\phi,v)$ must be in this set.
    
    Write $\rho = \rho_2$.
    Note that $\omega_q = -1$ for $q=2$.
    This implies $\Psi_2(z) = \phi(z)\phi(-z) = \rho(z^2)$ and $c(z) = \phi(-z)$ with coefficients $c_k = (-1)^{k+1}\phi_k$ s.t. $\gamma_u(h) = v\sum_{k}\phi_k\phi_{k+2h}$.
    Also, under $q=2$, $[Q]_2(z) = \frac{1}{2}(Q(z)+(Q-z))$ keeps only the even powers of $Q$.
    At the complex number $\ii$, we have $\ii^{-1} = -\ii$ and $\ii^2 = -1$ and therefore $[\Phi]_2(\ii)= \rho(-1)$ because 
    $\Phi(\ii) = \phi(\ii)\phi(-\ii) = \rho(-1)$ and likewise $\Phi(-\ii) = \rho(-1)$.
    By \cref{prop_ARpq_stock_observeddynamics_ID}, we have $\tilde\rho = \rho$ and $\Gamma_{\tilde u} = \Gamma_u$.

    %
    Note that $[\Phi]_2(z) = \Gamma_u(z^2)/v$.\footnote{
        To see this, write $\Phi(z) = \phi(z)\phi(z^{-1}) = \sum_{j,k}\phi_j\phi_k z^{j-k}$. 
        For $h\ge 0$, the coefficient of $z^{2h}$ is $\sum_k \phi_{k+2h}\phi_k$, which equals $\gamma_u(h)/v$.
        The same coefficient is obtained for $z^{2h}$ with $h \le 0$. 
        Therefore, the even powers-part of $\Phi$ is $\sum_{h}(\gamma_u(h)/v)z^{2h} = \Gamma_u(z^2)/v$.
    } 
    Evaluating this expression at $z=\ii$ and rearranging yields $v = \Gamma_u(-1)/\rho(-1)$.
    Since $\phi$ has real coefficients, $\phi(-\ii) = \overline{\phi(\ii)}$, so $\rho(-1) = \phi(\ii)\phi(-\ii) = |\phi(\ii)|^2 \ge 0$, and it is non-zero because by \cref{ass_ARpq_ID_stationarity} all zeros of $\phi$ satisfy $|z|>1$ while $|\ii|=1$. 
    This identity holds at any point in $\Theta_p$.
    Since $\tilde\rho = \rho$ and $\Gamma_{\tilde u} = \Gamma_u$, we must also have $\tilde v = v$.

    Next, I show that either $\tilde\Phi(z) = \Phi(z) \; \forall \; z$ or $\tilde\Phi(z) = \Phi(-z) \; \forall \; z$. 
    For this purpose, first define $P_1(z) = 2\Gamma_u(z^2)/v$ and $P_2(z) = \rho(z^2)\rho(z^{-2})$,
    note that they are point-identified,
    and note that $P_1(z) = \tilde\Phi(z) + \tilde\Phi(-z)$ and $P_2(z) = \tilde\Phi(z) \tilde\Phi(-z)$.
    Since both $P_1$ and $P_2$ are point-identified, these expressions hold likewise with and without tildes on $\Phi$, i.e. both for the $\Phi$ implied by the true parameter and that implied by any observationally equivalent $(\tilde\phi,\tilde v)$.
    For any scalar unknown $X$,
    $$ (X - \Phi(z))(X-\Phi(-z)) = X^2 - P_1(z)X + P_2(z) = 0 \; . $$
    Hence, for any fixed $z$, the two numbers $\Phi(z)$ and $\Phi(-z)$ are
    the two roots of this quadratic equation.
    Now, define $D_1(z) = \tilde\Phi(z) - \Phi(z)$ and $D_2(z) = \tilde\Phi(z) - \Phi(-z)$, 
    and note that 
    \begin{align*}
        D_1(z) D_2(z)
        &= \tilde\Phi(z)^2 - \tilde\Phi(z)\sbr{ \Phi(z)+\Phi(-z) } + \Phi(z)\Phi(-z) \\
        &= \tilde\Phi(z)^2 - \tilde\Phi(z) P_1(z) + P_2(z) \\
        &= \tilde\Phi(z)^2 - \tilde\Phi(z)\sbr{ \tilde\Phi(z)+\tilde\Phi(-z) }
        + \tilde\Phi(z)\tilde\Phi(-z) = 0  \; ,
    \end{align*}
    for any $z\neq0$,
    which implies that $\tilde\Phi(z)\in\{\Phi(z),\Phi(-z)\}$.
    This pointwise statement leaves open that the candidate agrees with
    $\Phi(z)$ for some values of $z$ and with $\Phi(-z)$ for others.
    To rule such switching, I show that one of $D_1$ or $D_2$ must be the zero polynomial, implying $\tilde\Phi(z)=\Phi(z) \; \forall \; z$ or
    $\tilde\Phi(z)=\Phi(-z) \; \forall \; z$.
    Suppose, by contradiction, that neither $D_1$ nor $D_2$ is the zero function.
    A Laurent polynomial that is not identically zero and has lowest power $-p$ can be written as $z^{-p}$ times a non-zero ordinary polynomial and therefore has only finitely many zeros in $\complexnumbers\setminus\{0\}$. 
    Hence, for all $z$ outside a finite set, we have $D_1(z)\neq0$ and $D_2(z)\neq0$ and therefore $D_1(z)D_2(z)\neq0$, which contradicts $D_1(z)D_2(z)=0 \; \forall \; z$.

    If $\tilde\Phi(z) = \Phi(z) \; \forall \; z$, then \cref{lemma_phiCoincidesIfPhiCoincides} (with $\hat\phi = \phi$) gives $\tilde\phi = \phi$.
    If instead $\tilde\Phi(z) = \Phi(-z) \; \forall \; z$, then we can write
    \[
        \Phi(-z) = \phi(-z) \phi(-z^{-1}) = \phi^-(z) \phi^-(z^{-1}) \; ,
    \]
    and so \cref{lemma_phiCoincidesIfPhiCoincides} (with $\hat\phi = \phi^-$) gives $\tilde\phi = \phi^-$. 
    Hence $(\tilde\phi,\tilde v) \in \{(\phi,v),(\phi^-,v)\}$.
    $\blacksquare$
}


\myproofof{prop_ARpq_stock_ID_realroots}{
    By \cref{prop_ARpq_stock_ID_generalcase}, any observationally equivalent $(\tilde\phi,\tilde v)\in \Theta_p$ satisfies $\tilde\lambda_i^q = \lambda_i^q$ and $K_i(\tilde\phi,\tilde v) = K_i(\phi,v)$ for $i=1:p$.
    By \cref{lemma_aliasingbound}, then, and since all $\lambda_i$ are real, we must have $\tilde\lambda_i = \pm \lambda_i$ and, therefore, also $\tilde\lambda_i^2 = \lambda_i^2$ for $i=1:p$.

    Consider first $q$ even.
    If $(\phi,v)$ satisfies \cref{ass_ARpq_ID_distincteigvalqthpowers} at some even $q$, it also satisfies it at $q=2$.
    Hence, we can apply \cref{prop_ARpq_stock_ID_generalcase} to conclude that this $(\tilde\phi,\tilde v)$ is observationally equivalent to $(\phi,v)$ also under the $q=2$-sampling scheme.
    By \cref{prop_ARp2_stock_ID}, then, $(\tilde\phi,\tilde v) \in \{(\phi,v),(\phi^-,v) \}$.
    Conversely, under $q$ even, $(\phi^-,v)$ is obervationally equivalent to $(\phi,v)$ by \cref{prop_ARpq_stock_ID_qEven_reflectionequiv}.

    Consider now $q$ odd.
    By \cref{lemma_aliasingbound}, since all $\lambda_i$ are real and $q$ is odd, we must have $\tilde\lambda_i = \lambda_i$ for $i=1:p$.
    By \cref{lemma_rootsandpolynomials}, this root multiset translates into a unique real polynomial with degree $p$ and constant term 1, hence $\tilde\phi=\phi$.
    Finally, \cref{prop_ARpq_stock_ID_generalcase} gives $K_i(\phi,\tilde v) = K_i(\phi,v)$, and since $K_i$ is proportional to $v$ with non-zero factor, we must have $\tilde v = v$.
    $\blacksquare$
}

\propositionmarker{prop_ARpq_flow_observeddynamics_ID}

\myproofof{prop_ARpq_flow_observeddynamics_ID}{
    Write $\mu_i = \lambda_i^q$.
    We have $\gamma_y(h) = \gamma_X(qh)$ and $\gamma_X(h) = \sum_{|m|\leq q-1}(q-|m|)\gamma_x(h+m)$. 
    For $h \geq q-1$, we have $h+m \geq 0$ for all $|m| \leq q-1$, so we can apply \cref{lemma_ACF_ARp_truevalue} to get $\gamma_X(h) = \sum_i K_i \sbr{ \sum_{|m|\leq q-1}(q-|m|)\lambda_i^m }\lambda_i^h = \sum_i K_iS_q(\lambda_i)\lambda_i^h$ for $h \geq q-1$.
    This yields the expression for $\gamma_y(h)=\gamma_X(qh)$ for $h\geq 1$, as then $qh\geq q>q-1$.
    We have $K_i^X \neq 0$ because $K_i \neq 0$ and
    $S_q(\lambda_i) = s_q(\lambda_i)s_q(1/\lambda_i) \neq 0$.

    The candidate must have effective order $\tilde p \geq 1$.\footnote{
        If $\tilde p = 0$, then $\gamma_{\tilde x}(h) = \tilde v\mathbf{1}\{h=0\}$ and
        $\gamma_{\tilde y}(h) = \gamma_{\tilde X}(qh) = (q-qh) \tilde v\mathbf{1}\{qh=0, qh \leq q-1\} = 0$ for $h \geq 1$, so $\sum_i K_i^X \mu_i^h = 0$ for $h \geq 1$ and \cref{lemma_unique_expopolysums} yields $K_i^X = 0$ for all $i$, a contradiction. 
    } 
    So by \cref{lemma_ACF_ARp_candidate},
    $$ \gamma_{\tilde y}(h) = \gamma_{\tilde X}(qh) = \sum_{|m|\leq q-1}(q-|m|)\gamma_{\tilde x}(qh+m) = \sum_{i=1}^{\tilde d} \tilde P_i^X(qh)\br{ \tilde\lambda_i^q }^h \; \; \text{for} \; h \geq 1 \;, $$
    where $\tilde P_i^X(h) = \sum_{|m|\leq q-1}(q-|m|)\tilde\lambda_i^{m}\tilde P_i(h+m)$.
    The merging argument in the proof of \cref{prop_ARpq_stock_observeddynamics_ID} applies one-to-one and yields $\{\tilde\lambda_i^q\}_{i=1}^p = \{\lambda_i^q\}_{i=1}^p$.
    By \cref{lemma_aliasingbound}, we can uniquely index the candidate roots s.t. $\tilde\lambda_i^q = \lambda_i^q$ for $i=1:p$.
    Applying \cref{lemma_ACF_ARp_truevalue} to $(\tilde\phi,\tilde v)$, we can see that (with re-indexed roots) $K_i^X(\tilde\phi,\tilde v) = K_i^X(\phi,v)$ for each $i$.
    Just as in the proof of \cref{prop_ARpq_stock_observeddynamics_ID}, $\tilde\rho_q = \rho_q$ and $\Gamma_{\tilde u} = \Gamma_u$ then follows by \cref{lemma_PsiqRewriting} and using \cref{prop_ARpq_flow_observeddynamics}.
    $\blacksquare$
}

\propositionmarker{prop_ARpq_flow_ID}

\propositionpartmarker{prop_ARpq_flow_ID_generalcase}

\propositionpartmarker{prop_ARp2_flow_ID}

\propositionpartmarker{prop_ARpq_flow_ID_qEven_reflectionnonequiv}

\propositionpartmarker{prop_ARpq_flow_ID_realrootsOddq}

\myproofof{prop_ARpq_flow_ID_generalcase}{
    Any observationally equivalent $(\tilde\phi,\tilde v)$ has $\gamma_{\tilde y}(0) = \gamma_y(0)$,
    and by \cref{prop_ARpq_flow_observeddynamics_ID}, it also satisfies the other two conditions.
    Conversely, let $(\tilde\phi,\tilde v)$ be in the stated set.
    Then, by \cref{lemma_aliasingbound}, it satisfies \cref{ass_ARpq_ID_actualorderisp,ass_ARpq_ID_distincteigvalqthpowers}, 
    and therefore has non-zero and distinct roots.
    Using \cref{lemma_ACF_ARp_truevalue} and repeating the derivation in the first paragraph of the proof of \cref{prop_ARpq_flow_observeddynamics_ID} with $(\tilde\phi,\tilde v)$ in place of $(\phi,v)$ yields
    $\gamma_{\tilde y}(h) = \sum_{i=1}^p K_i^X(\tilde\phi,\tilde v)\br{ \tilde\lambda_i^q }^h$ for $h \geq 1$. 
    Using $\tilde\lambda_i^q = \lambda_i^q$ and $K_i^X(\tilde\phi,\tilde v) = K_i^X(\phi,v)$,
    $$ \gamma_{\tilde y}(h) = \sum_{i=1}^p K_i^X(\phi,v) \br{ \lambda_i^q }^h = \gamma_y(h) \quad \text{for } h \geq 1\;, $$
    while $\gamma_{\tilde y}(0) = \gamma_y(0)$ holds by assumption.
    Because $y$ is Normal --- implied by \cref{ass_ARpq_ID_normality} --- and has mean zero,
    this corresponds to observational equivalence.
    $\blacksquare$
}


\myproofof{prop_ARp2_flow_ID}{
    Write $\rho = \rho_2$, $s(z) = s_2(z) = (1+z)$ and $S(z)=S_2(z)=s(z)s(z^{-1})=2+z+z^{-1}$.
    Recall that $\wq=-1$ for $q=2$ s.t. $\Psi_2(z)=\phi(z)\phi(-z)=\rho(z^2)$ and $c(z) = \phi(-z)$, as in the proof of \cref{prop_ARp2_stock_ID}.
    In addition, note that $S(\ii)=S(-\ii)=2$, $S(1)=4$ and $S(-1)=0$ and $S(z)S(-z)=2-z^2-z^{-1}$.
    Define the symmetric Laurent polynomial $\Xi(z) = \Phi(-z)S(z)$, 
    and note that $\Xi(z)\Xi(-z) = \rho(z^2)\rho(z^{-2})(2-z^2-z^{-2})$ and $[\Xi]_2(\ii) = \frac{1}{2}(\Xi(\ii)+\Xi(-\ii)) = \frac{1}{2}(\Phi(-\ii)S(\ii)+\Phi(\ii)S(-\ii)) = \Phi(\ii)+\Phi(-\ii)= 2\rho(-1)$.
    By \cref{prop_ARpq_flow_observeddynamics_ID}, we have $\tilde\rho = \rho$ and $\Gamma_{\tilde u} = \Gamma_u$.

    Note that $[\Xi]_2(z) = \Gamma_u(z^2)/v$.\footnote{
        By the proof of \cref{prop_ARpq_flow_observeddynamics}, $u_t = d(L)e_{2t}$ with $d(z) = \phi(-z)s(z)$, so $\gamma_u(h)/v = \sum_k d_k d_{k+2h}$ is the coefficient of $z^{2h}$ in $d(z)d(z^{-1}) = \phi(-z)\phi(-z^{-1}) s(z)s(z^{-1}) = \Phi(-z)S(z) = \Xi(z)$, for $h \ge 0$ and likewise for $h \le 0$. Therefore, the even powers-part of $\Xi$ is $\sum_h(\gamma_u(h)/v)z^{2h} = \Gamma_u(z^2)/v$.} 
    Evaluating this expression at $z = \ii$ and rearranging yields $v = \Gamma_u(-1)/\br{2\rho(-1)}$, where $\rho(-1) = |\phi(\ii)|^2 > 0$ as in the proof of \cref{prop_ARp2_stock_ID}. 
    This identity holds at any point in $\Theta_p$. 
    Since $\tilde\rho = \rho$ and $\Gamma_{\tilde u} = \Gamma_u$, we must also have $\tilde v = v$.

    Next, define $P_1(z) = 2\Gamma_u(z^2)/v$ and $P_2(z) = \rho(z^2)\rho(z^{-2})(2 - z^2 - z^{-2})$, note that they are point-identified, and note that $P_1(z) = \Xi(z) + \Xi(-z)$ and $P_2(z) = \Xi(z) \Xi(-z)$. Since both $P_1$ and $P_2$ are point-identified, these expressions hold likewise with and without tildes on $\Xi$. Define $D_1(z) = \tilde\Xi(z) - \Xi(z)$ and $D_2(z) = \tilde\Xi(z) - \Xi(-z)$. 
    By the same argument as in the proof of \cref{prop_ARp2_stock_ID} -- with $\Xi$ replacing $\Phi$ --, we get $\tilde\Xi(z) = \Xi(z) \;\forall\; z$ or $\tilde\Xi(z) = \Xi(-z) \;\forall\; z$.

    In fact, we must have $\tilde\Xi(z) = \Xi(z) \;\forall\; z$.
    Suppose instead $\tilde\Xi(z) = \Xi(-z)$ for all $z$, and evaluate this at $z = -1$. 
    On the left side, $\tilde\Xi(-1) = \tilde\Phi(1) S(-1) = 0$, since $S(-1) = 0$. 
    On the right side, $\Xi(1) = \Phi(-1) S(1) = 4 \phi(-1)^2$, which is non-zero by \cref{ass_ARpq_ID_stationarity}.

    Finally, $\tilde\Xi = \Xi$ means $\br{\tilde\Phi(-z) - \Phi(-z)}S(z) = 0$ for all $z \neq 0$. Since $S \neq 0$, we must have $\tilde\Phi = \Phi$. 
    By \cref{lemma_phiCoincidesIfPhiCoincides}, then, $\tilde\phi = \phi$.
    $\blacksquare$
}


\myproofof{prop_ARpq_flow_ID_qEven_reflectionnonequiv}{
    Suppose, by contradiction, $(\phi^-,\tilde v)$ is observationally equivalent to $(\phi,v)$.
    Note that $\phi^-$ has roots $\tilde\lambda_i = -\lambda_i$.
    By \cref{prop_ARpq_flow_observeddynamics_ID}, we must have $K_i(\phi^{-},\tilde v)S_q(-\lambda_i) = K_i(\phi,v)S_q(\lambda_i)$ for $i=1:p$.
    Note that $K_i(\phi^-,\tilde v) = (\tilde v/v)K_i(\phi,v)$ because $\phi^{-}$ has roots $\{-\lambda_i\}_{i=1}^p$.
    Therefore, 
    $$ \frac{\tilde v}{v} = \frac{S_q(\lambda_i)}{S_q(-\lambda_i)} = \frac{(1+\lambda_i)(1+\lambda_i^{-1})}{(1-\lambda_i)(1-\lambda_i^{-1})} = -\left(\frac{1+\lambda_i}{1-\lambda_i}\right)^{\!2} \; , $$
    using $S_q(z) = s_q(z)s_q(z^{-1})$ and $s_q(z) = (1-z^{q})/(1-z)$ for $z\neq 1$ as well as the fact that $z^q = (-z)^q$ for $q$ even.
    Let $w_i = (1+\lambda_i)/(1-\lambda_i)$ and note that $\mathrm{Re}(w_i) > 0$.\footnote{
        We have $w_i = (1+\lambda_i)(1-\overline{\lambda_i})/|1-\lambda_i|^2 = (1-|\lambda_i|^2 + \lambda_i-\overline{\lambda_i})/|1-\lambda_i|^2$.
        Since $\lambda_i-\overline{\lambda_i}$ has no real part, $\mathrm{Re}(w_i) = (1-|\lambda_i|^2)/|1-\lambda_i|^2 > 0$ since $|\lambda_i|<1$ by \cref{ass_ARpq_ID_stationarity}.
    } 
    Since $\tilde v/v > 0$, we need $-w_i^2 > 0$, i.e.\ $w_i^2$ must be a negative real number. 
    Writing $w_i = a + b\ii$, we have $w_i^2 = (a^2 - b^2) + 2ab\ii$, which is a negative real number iff $a = 0$ and $b \neq 0$, which contradicts $\mathrm{Re}(w_i) > 0$.
    $\blacksquare$
}


\myproofof{prop_ARpq_flow_ID_realrootsOddq}{
    By \cref{prop_ARpq_flow_observeddynamics_ID}, any observationally equivalent $(\tilde\phi,\tilde v)$ satisfies $\tilde\lambda_i^q = \lambda_i^q$ for $i=1:p$.
    By \cref{lemma_aliasingbound}, then, and since all $\lambda_i$ are real, we must have $\tilde\lambda_i = \pm \lambda_i$ for $i=1:p$.
    Since $q$ is odd, we have $\tilde\lambda_i = \lambda_i$ for $i=1:p$.
    By \cref{lemma_rootsandpolynomials}, this root multiset translates into a unique real polynomial with degree $p$ and constant term 1, hence $\tilde\phi=\phi$.

    It remains to show that $\tilde v = v$. 
    \cref{prop_ARpq_flow_observeddynamics_ID} also gives $K_i^X(\tilde\phi,\tilde v) = K_i^X(\phi,v)$ for $i=1:p$, 
    which is $K_i(\phi,\tilde v) S_q(\lambda_i) = K_i(\phi,v) S_q(\lambda_i)$ under $\tilde\phi=\phi$ and $\tilde\lambda_i = \lambda_i$ for $i = 1:p$.
    Since $S_q(\lambda_i) \neq 0$, we get $K_i(\phi,\tilde v) = K_i(\phi,v)$. 
    By the formula in \cref{lemma_ACF_ARp_truevalue}, $K_i(\phi,\tilde v)/K_i(\phi,v) = \tilde v/v$, and hence $\tilde v = v$.
    $\blacksquare$
}

\propositionmarker{prop_ARpq_ID_computationofIDset}

\myproofof{prop_ARpq_ID_computationofIDset}{
    Consider first the stock variable-case.
    \cref{prop_ARpq_stock_ID_generalcase} characterizes $\calI^{s,1}_{p,q}(\phi,v)$: 
    each point in it satisfies $\tilde\lambda_i^q = \lambda_i^q$ and $K_i(\tilde\phi,\tilde v) = K_i(\phi, v)$ for $i=1:p$.
    By statement (v) of \cref{lemma_aliasingbound}, there are $G$ different $\tilde\phi$ s.t. $\tilde\lambda_i^q = \lambda_i^q$ for $i=1:p$, and they can be computed using the expressions in statements (iii) and (iv) of \cref{lemma_aliasingbound}.
    Note that each such $\tilde\phi$ satisfies \cref{ass_ARpq_ID_stationarity} by \cref{lemma_aliasingbound}, so $(\tilde\phi,\tilde v)\in\Theta_p$ for any $\tilde v>0$, and $D_i(\tilde\phi)$ is well-defined and non-zero for $i=1:p$, by the same argument as in the proof of \cref{lemma_ACF_ARp_truevalue}.
    Then, note that $K_i(\tilde\phi,\tilde v) = \tilde v/D_i(\tilde\phi)$ s.t. $K_i$-matching for $i=1:p$ corresponds to $D_i(\tilde\phi)/D_i(\phi) = \tilde v/v$ for $i=1:p$.
    This also shows that each such $\tilde\phi$ is consistent with at most one $\tilde v$, which means that there are at most $G$ points in $\calI^{s,1}_{p,q}(\phi,v)$.

    Now suppose $y_t$ is a flow variable.
    \cref{prop_ARpq_flow_ID_generalcase} characterizes $\calI^{f,1}_{p,q}(\phi,v)$: each point in it satisfies $\tilde\lambda_i^q = \lambda_i^q$ and $K^X_i(\tilde\phi,\tilde v) = K^X_i(\phi, v)$ for $i=1:p$ as well as $\gamma_{\tilde y}(0) = \gamma_y(0)$.
    As in the stock variable-case, \cref{lemma_aliasingbound} shows that there are $G$ different $\tilde\phi$ s.t. $\tilde\lambda_i^q = \lambda_i^q$ for $i=1:p$ and how to compute them.
    Then, note that $K_i^X(\tilde\phi,\tilde v) = \tilde v S_q(\tilde\lambda_i)/D_i(\tilde\phi)$ with $S_q(\tilde\lambda_i) \neq 0$, so $K_i^X$-matching for $i=1:p$ corresponds to $\sbr{D_i(\tilde\lambda) S_q(\lambda_i)} / \sbr{D_i(\lambda) S_q(\tilde\lambda_i)} = \tilde v/v$ for $i=1:p$.
    Again, this also shows that each such $\tilde\phi$ is consistent with at most one $\tilde v$ and proves the count for $\calI^{f,1}_{p,q}(\phi,v)$.
    The second check is the remaining condition $\gamma_{\tilde y}(0) = \gamma_y(0)$ of \cref{prop_ARpq_flow_ID_generalcase}, evaluated at $\tilde v = \kappa^{(g)}v$. 
    $\blacksquare$
}


\section{Estimation}
\label{appsec_est}

\paragraph*{Estimation Precision}

For the stock case with $p=1$, the observed process is an AR($1$):
$$ y_t=\psi y_{t-1}+\varepsilon_t \; , \quad \varepsilon_t = \sum_{l=0}^{q-1}\phi^l e_{tq-l} \sim WN(0,v_q) \; , \quad  \psi=\phi^{q} \; , \quad v_q=v\,\frac{1-\phi^{2q}}{1-\phi^{2}} \; , $$
for which $\sqrt{T}(\hat\psi-\psi) \convdist N(0,1-\psi^2)$, independently of $\hat{v}_q$ \citep[Propositions 3.1 and 3.4]{Luetkepohl2005}; see also \citet[Example 8.8.1]{BrockwellDavis1991} and Eq. (21) in \citet{NijmanPalm1990b}.

%
Since $\psi$ depends on $\phi$ alone, we have $\hat\psi = \hat\phi^{q}$.
Provided $\phi$ is identified, the delta method yields $\sqrt{T}(\hat\phi-\phi) \convdist N\br{0,(\partial\psi/\partial\phi)^{-2}(1-\psi^2)}$.
Writing $\calI_{\phi\phi}(q)$ for the inverse of this asymptotic variance, the per-observation information about $\phi$ is
\begin{align*} 
  \calI_{\phi\phi}(q) = 
  \Bigl(\frac{\partial\psi}{\partial\phi}\Bigr)^{\!2}\frac{1}{1-\psi^{2}} = 
  \frac{q^{2}\phi^{2q-2}}{1-\phi^{2q}} \; .
\end{align*}
At $q=1$ this returns the well-known $\br{1-\phi^{2}}^{-1}$.
The same expression is the first term in Eq. (B.3) of \citet{NijmanPalm1990b}, and their Eq. (22) equals $q/\calI_{\phi\phi}(q)$: their sample size counts high-frequency periods, i.e. equals $q$ times my number of observations.
At $q=2$, $\calI_{\phi\phi}(1)/\sbr{\calI_{\phi\phi}(2)/2} = (1+\phi^2)/(2\phi^2)$ is the relative efficiency of the complete-data estimator in \citet[p.~1427]{PalmNijman1984}.

Differentiating $\calI_{\phi\phi}(q)$ gives
\[
  \frac{d}{dq}\log\calI_{\phi\phi}
  =\frac{2}{q}+\frac{2\log|\phi|}{1-\phi^{2q}} \; .
\]
This shows that $\calI_{\phi\phi}(q)$ does not necessarily always decrease with $q$. Instead, it is maximized at 
$q^{*} \approx - 0.797 / \log(|\phi|)$.\footnote{
    The FOC is $1/q = -\log(|\phi|)/(1-\phi^{2q})$.
    Substituting $u=-2q\log(|\phi|)$ so that $\phi^{2q}=e^{-u}$, the condition becomes $2/u=1/(1-e^{-u})$, i.e. $u=2-2e^{-u}$, whose positive root is $u^{*}\approx1.5936$.
    Using this, we can solve for $q^{*} = u^{*}/(-2\log(|\phi|))$.
}

\end{appendix}

\end{document}